\documentclass{JHEP3}
\usepackage{epsf}
\usepackage{amsmath}
\usepackage{amsfonts}

\newcommand{\onefigure}[2]{\begin{figure}[htbp]
         \caption{#2\label{#1}(#1)}
         \end{figure}}

\renewcommand{\onefigure}[2]{\begin{figure}[htbp]
         \begin{center}\leavevmode\epsfbox{#1.eps}\end{center}
         \caption{#2\label{#1}}
         \end{figure}}
\newcommand{\comment}[1]{}

\newcommand{\figref}[1]{Fig.~\protect\ref{#1}}

\def\bbbz{{\sf Z\!\!\!Z}}

\def\sl2z{SL(2,\bbbz)}
\newcommand{\be}{\begin{equation}}
\newcommand{\ee}{\end{equation}}
\newcommand{\bea}{\begin{eqnarray}}
\newcommand{\eea}{\end{eqnarray}}
\newcommand{\nn}{\nonumber}
\newcommand{\unit}{1\!\!1}

\def\bbbz{{\sf Z\!\!\!Z}}
\def\sl2z{SL(2,\bbbz)}

\def\z0{{\bf z_0}}

\def\IC{\mathbb{C}}
\def\IZ{\mathbb{Z}}
\def\IF{\mathbb{F}}

\def\beq{\begin{equation}}
\def\eeq{\end{equation}}
\def\beqa{\begin{eqnarray}}
\def\eeqa{\end{eqnarray}}
\def\beqan{\begin{eqnarray*}}
\def\eeqan{\end{eqnarray*}}

\preprint{MIT-CTP-3264\\ UTTG-19-01 \\ {\tt hep-th/0206152}}
\title{Quiver Theories, Soliton Spectra and Picard-Lefschetz transformations}
\author{
Bo Feng$^{~1}$, Amihay Hanany$^{~1}$, Yang Hui He$^{~1}$, Amer
Iqbal$^{~2}$ \footnote{ Research supported in part by the Reed
Fund Award, the CTP and the LNS of MIT and the U.S. Department of
Energy under cooperative research agreement \# DE-FC02-94ER40818
as well as the NSF under Grant No. 0071512. A.~H.~ is also
supported by a DOE OJI Award.}
\\
$^{1}$ Center for Theoretical Physics\\
Department of Physics, MIT\\
Cambridge, MA 02139.
\vskip 0.5cm
$^{2}$ Theory Group, Department of Physics\\
University of Texas at Austin,\\
Austin, TX 78712.\\
\email{fengb, hanany, yhe@ctp.mit.edu; iqbal@physics.utexas.edu}
}
\abstract{ Quiver theories arising on D3-branes at orbifold and
del Pezzo singularities are studied using mirror symmetry. We show
that the quivers for the orbifold theories are given by the
soliton spectrum of massive $2d$ ${\cal N}$=2 theory with weighted
projective spaces as target. For the theories obtained from the
del Pezzo singularities we show that the geometry of the mirror
manifold gives quiver theories related to each other by
Picard-Lefschetz transformations, a subset of which are simple
Seiberg duals. We also address how one indeed derives Seiberg
duality on the matter content from such geometrical transitions
and how one could go beyond and obtain certain
 ``fractional Seiberg duals.''  Moreover, from the mirror geometry for
the del Pezzos arise certain Diophantine equations which classify
all quivers related by Picard-Lefschetz.  Some of these
Diophantine equations can also be obtained from the classification
results of Cecotti-Vafa for the $ 2d$ ${\cal N}=2$ theories.
\keywords{D-brane Probes, del Pezzo Surfaces, Mirror Symmetry,
Seiberg Duality, Soliton Spectra} }
\begin{document}


\section{Introduction}

The technology of D3-branes probing singularities
as a method of establishing classes of gauge theories in four dimensions
is by now a well-establish subject. Indeed we are interested in
local algebraic models of non-compact Calabi-Yau threefolds that give
${\cal N}=1$ supersymmetric gauge theories.
In addition to the orbifolds pioneered by Douglas-Moore
and followups \cite{orbifold}, toric singularities have also been
extensively investigated \cite{phases,torichist,toric,toric2}.

Attention has been paid of late to del Pezzo surfaces
\cite{mystery,toric,HI,HaIq}. Indeed with the r\^{o}le of mirror
symmetry \cite{HV,HIV} in the geometrisation of ${\cal N}=1$
dualities \cite{CKV,CIV,CFIKV,OV2,DOT}, D3-branes probing the cone
over del Pezzo surfaces as well as the mirror perspective of
D6-branes wrapping special Lagrangian three-cycles have been
increasingly important. An intriguing matter has been the
realisation of Seiberg's duality in terms of what has been called
``Toric Duality" \cite{toric,toric2,chris,seiberg,multi}.

Treading upon this path, the quiver theories we are interested in are
${\cal N}=1$, supersymmetric $D=4$ gauge theories arising on the
worldvolume of D3-branes transverse to a Calabi-Yau threefold with del
Pezzo singularity in the Type IIB background. If the singularity is
not an orbifold singularity it is difficult to obtain the information
about the gauge groups and the matter content. However, it was shown
in \cite{HaIq} that mirror symmetry provides a powerful tool in
determining the gauge groups and the quiver diagrams representing the
matter. In \cite{CFIKV} mirror symmetry was used to engineer Seiberg
dual theories arising on the toric del Pezzo singularities and a
conjecture was given for calculating the superpotential. It was shown
that under certain Picard-Lefshetz transformation the
superpotential transforms as expected from Seiberg duality. Also
in the case of $\mathbb{P}^{1}\times \mathbb{P}^{1}$ a duality cascade
was engineered using the results of \cite{HIV}.

In this paper we continue to study the gauge theories arising on
the D3-branes from the mirror symmetry perspective as D6-branes
wrapped on 3-cycles in Type IIA. We study the case of  del Pezzos
in more detail by giving an exceptional collection forming a helix
on the del Pezzo surface and show that this exceptional collection
gives the correct Ramond charges for the massive theory with del
Pezzo surfaces as the target space. The exceptional collections
also give the charges of the fractional branes in the Type IIB
description and therefore the constraint that we get correct
Ramond charges from a collection of bundles on the del Pezzo
surfaces gives us certain Diophantine equations which classifies
all quiver gauge theories related to each other by
Picard-Lefschetz transformations. We also obtain the same
Diophantine equations from the geometry of the Calabi-Yau mirror
to the local del Pezzos.

Besides del Pezzos, orbifold singularities provide an interesting
class of singularities of the Calabi-Yau threefolds.  These
orbifold singularities arise when a four cycle which is a weighted
projective space collapses. However, if we resolve the
singularities of the weighted projective space as well then the
singularity is produced by multiple four cycles collapsing. Mirror
symmetry is a powerful tool for studying the gauge theories
obtained from these singularities\footnote{Also for non-toric
singularities once the mirror manifold is determined.} and gives a
geometric interpretation to  Seiberg duality \cite{CFIKV,DOT}.

Under mirror symmetry D3-branes transverse to a non-compact Calabi-Yau
threefold $X$ become D6-branes wrapped on a $T^{3}$ in the mirror
Calabi-Yau $Y$ \cite{SYZ,HaIq,CFIKV}.  The homology class of this $T^{3}$ is
given by
\bea
\label{t3}
[T^{3}]=\sum_{a=1}^{K}n_{a}S_{a}\,,\,\,n_{a}\in \mathbb{Z}\,,
\eea
where $\{S_{1},\cdots, S_{K}\}$, which  form a basis of $H_{3}(Y,\mathbb{Z})$,
are three cycles
 topologically equivalent to $S^{3}$ and $n_a$ is the wrapping number of
cycle $S_a$.  The D6-brane wrapped on $T^{3}$ gives rise to a
${\cal N}=1$ $D=4$ theory with gauge group $G$ and quiver matrix
$Q$ given by \cite{HaIq,CFIKV} \bea
G=\prod_{a=1}^{K}U(n_{a})\,,\,\,\,Q_{ab}=S_{a}\cdot S_{b}\,. \eea
In the above equations we have assumed $n_{a}\geq 0$ which can
always be arranged by changing the orientation of the 3-cycles
$S_{a}$. The quiver matrix is just the intersection matrix of the
3-cycles. In terms of the fractional branes $\{{\cal
F}_{1},\cdots, {\cal F}_{K}\}$ (which are mirror to $S_{a}$) on
$X$, on Type IIB side, this is given by \cite{HI,MOY,HaIq,CFIKV} \bea \label{QAB}
Q_{ab}=\int_{X} \mbox{ch}({\cal F}_{a}\otimes {\cal
F}_{b}^{*})\mbox{Td}(X)\,. \eea The anomaly cancellation condition
is given by \bea \label{1.4}
0=\sum_{b=1}^{K}n_{b}Q_{ab}=\sum_{b=1}^{K}S_{a}\cdot
n_{b}S_{b}=S_{a}\cdot [T^{3}]\,,\,\forall a\,. \eea and the fact
that it is satisfied automatically follows from the geometry of
the mirror manifold  \cite{HaIq}. Equation (\ref{t3}) gives us a
particular solution to the anomaly cancellation condition, more
general solutions $\{n_{1},\cdots,n_{K}\}$ can also be found such
that $\sum_{i=1}^K n_{i}S_{i}$ is not topologically a $T^{3}$, but
still has $zero$ intersections with all $S_{a}$.

We can rephrase Equations (\ref{QAB}) and (\ref{1.4})
in the language of exceptional
collections of vector bundles (or sheaves) over the compact
divisor of the Calabi-Yau $X$. (q.v.~ \cite{HV,HIV,HaIq,CFIKV}). Given
an exceptional collection \bea \{F_{1},\cdots, F_{K}\} \eea such
that\footnote{$\mbox{ch}_{0}(V),ch_{1}(V), ch_{2}(V)$ are,
respectively, the rank, first Chern class and second Chern character
of $V$.} \bea \label{anomaly2} \sum_{a=1}^{K}n_{a}\mbox{ch}_{0}(F_{a})=0,\,\,
 \sum_{a=1}^{K}n_{a}\mbox{ch}_{1}(F_{a})=0,\,\,
 \sum_{a=1}^{K}n_{a}\mbox{ch}_{2}(F_{a})=-1,\,\,
\eea
we get an anomaly free gauge theory with gauge group and quiver given
by
\bea
G=\prod_{a=1}^{K}U(n_{a})\,,\,\,\,
Q_{ab}=\chi(F_{a},F_{b})-\chi(F_{a},F_{b}),
\label{index}
\eea
where $\chi(F_{i},F_{j}) := \sum_m (-1)^m \dim_{\IC} {\rm
Ext}^m(F_i,F_j)$.
For each subset of the exceptional collection $\{F_{\alpha_{1}},\cdots
F_{\alpha_{n}}\}$
there is a term in the superpotential \cite{CFIKV}(the bi-fundamental fields are
$X^{i}_{\alpha \beta}\in \mbox{Hom}(F_{\alpha}\otimes
F_{\beta}^{*})$):
\bea
a_{i_{1}i_{2}\cdots i_{n}}X^{i_{1}}_{\alpha_{1}\alpha_{2}}\cdots
X^{i_{n-1}}_{\alpha_{n-1}\alpha_{n}}X^{*\,i_{n}}_{\alpha_{1}\alpha_{n}},
\eea
where $X^{*\,i_{n}}_{\alpha_{1}\,\alpha_{n}}\in \mbox{Hom}
(F_{\alpha_{1}}\otimes F^{*}_{\alpha_{n}})$  and
$a_{i_{1}i_{2}\cdots i_{n}}$ are such that if $f^{i}_{\beta
\alpha}:F_{\alpha}\mapsto F_{\beta}$ then
\bea
f^{i_{n-1}}_{\alpha_{n}\alpha_{n-1}}\cdots
f^{i_{2}}_{\alpha_{3}\alpha_{2}}f^{i_{1}}_{\alpha_{2}\alpha_{1}}=a_{i_{1}i_{2}
\cdots
i_{n-1}i_{n}}f^{i_{n}}_{\alpha_{n}\alpha_{1}}\,.
\eea
In other words the terms of the superpotential come from non-zero loop
contractions in the quiver, where by contraction we mean composition
of maps.

Since $\{F_{1},\cdots F_{K}\}$ is an exceptional collection then
we can consider the left and the right mutations \cite{CFIKV},
with respect to the $l$-th node, \bea \{F_{l},F_{l+1}\}&\mapsto&
\{R_{l}F_{l+1},F_{l}\}\,,\\ \nn &\mapsto&  \{F_{l+1},
L_{l+1}F_{l}\}\,, \eea dictated by \bea
\mbox{ch}(R_{l}F_{l+1})=\mbox{ch}(F_{l+1})-\chi(F_{l},F_{l+1})
\mbox{ch}(F_{l})\,,\\ \nn
\mbox{ch}(L_{l+1}F_{l})=\mbox{ch}(F_{l})-\chi(F_{l},F_{l+1})
\mbox{ch}(F_{l+1})\,. \eea

Then, defining $\chi_{l}:=\chi(F_{l},F_{l+1})$, the changes on the
gauge group factors and the quiver diagram are:
\bea
\prod_{a}U(n_{a})&\mapsto&\prod_{a=1}^{l-1}U(n_{a})\,\,U(n_{l+1})U(n_{l}+\chi_{l}n_{l+1})
\prod_{a=l+2}^{N}U(n_{a})\,,\\
\nn
&\mapsto&
\prod_{a=1}^{l-1}U(n_{a})\,\,U(n_{l+1}+
\chi_{l}n_{l})U(n_{l})\prod_{a=l+2}^{N}U(n_{a})\,,\\
\eea
where it is easy to check that these new $n_a$ satisfy anomaly free
conditions (\ref{anomaly2}).
and
\bea
Q_{ab}&\mapsto& Q_{ab}\,,\,\,a,b\neq l,l+1\,,\\ \nn
Q_{a,l}&\mapsto&Q_{a,l+1}-\chi_{l}Q_{a,l}\,\\ \nn
        &\mapsto& Q_{a,l+1}\,,\\ \nn
Q_{a,l+1}&\mapsto& Q_{a,l}\,,\\ \nn
         &\mapsto& Q_{a,l}-\chi_{l}Q_{a,l+1}\,,\\\nn
Q_{l,l+1} &\mapsto& -Q_{l,l+1}\,,\\ \nn &\mapsto& -Q_{l,l+1}\,.
\eea Notice that if $n_{l}+\chi_{l}n_{l+1}< 0$, we should choose
the negative $R_l F_{l+1}$ as well as the $Q_{a,l+1}$ calculated
above. These mutations are also called Picard-Lefschetz
transformations, which we will discuss in detail in Section
\ref{five}. The field theory interpretation of these mutations is
nothing but a realization of Seiberg duality as will be discussed
throughout this paper.

The paper is organized as follows.  In Section \ref{N2} we briefly
review the classification of ${\cal N}=(2,2)$ theories, in
particular how one could obtain the quiver diagram of ${\cal N}=1$
probes on cones over del Pezzo as the soliton spectrum of these
massive theories in 2-dimensions.  Subsequently, we show how this
technique may be extended to the Abelian orbifold
$\mathbb{C}^3/\mathbb{Z}_{N}$ in Section \ref{Znorb}. We show how
we can use exceptional collections over weighted projective
spaces, as opposed to $\mathbb{P}^2$ and its blowups in the del
Pezzo case, to study the quiver theories. Explicit examples are
constructed for $\mathbb{C}^3/\mathbb{Z}_{5}$.  Then in Section
\ref{dPs}, we return to the case of the del Pezzos and study in
detail how we could wrap D6-branes on the mirror to obtain classes
of gauge theories related by Picard-Lefschetz monodromy. Therefrom
arise certain Diophantine equations which completely classifies
these theories.

We continue in this vein in Section \ref{five} where we show in
detail how one derives Seiberg duality rules for the matter
content from Picard-Lefschetz, characterized by ``$(p,q)$
7-brane'' moves and how one can go beyond and obtain ``fractional
Seiberg duality.''\footnote{By $(p,q)$ 7-brane here and in the
rest of the paper we just mean the marked point on the z-plane
over which the elliptic fiber has a degenerating $(p,q)$ cycle.}
In Section \ref{glosup} we briefly remark certain relations
between the superpotentials obtained in this setup and the global
isometries of the background geometry and also comment on the case
of $\mathbb{P}^2$, the zeroth del Pezzo, especially its
Diophantine equation, in some detail.  We end with Conclusions and
Prospects in Section \ref{conclud}.
\section{Classification of ${\cal N}=(2,2)$ two dimensional
theories and solitons} \label{N2}

In this section we collect few facts from the theory of massive
${\cal N}=(2,2)$ two dimensional theories and prepare their use
for quiver theories. We will see that the quiver diagram for the
four dimensional theories we are interested in are identified with
the soliton diagram of the massive two dimensional theory.

For a non-homogeneous superpotential $W$ of a massive LG theory,
the soliton spectrum is determined by the intersection number of
middle dimensional cycles in the geometry defined by \bea
W(x,y)=z\,. \label{super} \eea The middle dimensional cycles which
start at the critical points of the superpotential and project to
straight lines in the z-plane are the D-branes of the massive
theory \cite{HIV}. The intersection number of these middle
dimensional cycles calculates the Witten index in a sector in
which strings are stretched between the two D-branes given by the
cycles.

In \cite{HIV} it was shown that the intersection numbers of three
cycles in the mirror CY manifold, $Y$, give the soliton numbers of
the massive two dimensional theory with toric del Pezzo as the
target space. We will see that the geometry of the mirror CY $Y$
is completely captured by the four dimensional non-compact surface
defined by Equation (\ref{super}) for an appropriate $W$ and
therefore the quiver diagram, which is obtained from the
intersection number of three cycles, is identified with the
soliton diagram of the corresponding massive theory.

 From the classification results of \cite{CV} we know that an
arbitrary soliton diagram does not necessarily correspond to a
massive theory as the soliton spectrum is related to the Ramond
charges.  Let $A$ be an upper triangular matrix such that \bea
A_{ab}&=&0\,,a \leq b\,,\\ \nn A_{ab}&=&\mu_{ab}\,,a>b\,, \eea
where $\mu_{ab}$ is the number of solitons between the $a$-th and
the $b$-th vacua. The eigenvalues $\lambda_{a}$ of the matrix \bea
H=(1-A)(1-A)^{-T}\,, \eea are given by \bea \lambda_{a}=e^{2\pi i
q^{R}_{a}}\, \eea and thus are all phases. This follows from the
fact that the matrix $H$ is the monodromy matrix of the D-branes
of the massive theory as the massive superpotential $W$ goes to
$e^{2\pi i}W$. A derivation of this result is given in Section 4
of \cite{HIV}. The integer part of the Ramond charge $q^{R}_{a}$
can also be calculated as discussed in \cite{CV}. The fact that
the eigenvalues must be phases implies that the characteristic
polynomial of the matrix $H$ \bea P(z)=\mbox{det}(z-H) \eea is a
product of cyclotomic polynomials.

In the case that the target space is a compact K\"ahler manifold
of complex dimension $n$ which satisfies the condition on the Hodge numbers
$h^{p,q}=h^{p,p}\delta_{p,q}$, the Ramond charges are given by
$p-\frac{n}{2}$, each with multiplicity $h^{p,p}$. Specializing to
the case $n=2$, we find that for all del Pezzo surfaces
$\mathbb{B}_{k=0,\ldots,8}$ (with $h^{0,0}=h^{2,2}=1, h^{1,1}=k+1)$ the
eigenvalues are equal to one, since the charges are integral, and
thus the characteristic polynomial is \bea \label{charpoly}
P_{k}(z)=(z-1)^{k+3}\,,\,\, {\rm for} \,\,\mathbb{B}_{k}\,, \eea
where $k+3=\sum_{p}h^{p,p}(\mathbb{B}_{k}) = \chi(\mathbb{B}_{k})$.

As an example consider the case of $\mathbb{B}_{3}$. The quivers
for this case are given in \cite{toric,toric2,HaIq,seiberg,chris}
and all of them are related to each other by Picard-Lefschetz
transformation of three cycles. Consider case (IV) of
\cite{seiberg} (also case (IV) of \cite{chris}): \bea
A^{IV}=\begin{pmatrix}
0 & 0 & 0 & 0 & 0 & 0 \\ 0 & 0 & 0 & 0 & 0 & 0 \\0 & 0 & 0 & 0 & 0 & 0\\
1 & 1 & 1 & 0 & 0 & 0 \\1 & 1 & 1 & 0 & 0 & 0 \\ -2 & -2 & -2 & 3 & 3 & 0\\
\end{pmatrix}\,,
\eea the characteristic polynomial is given by \bea
P(z)=\mbox{det}(z-H)=(z-1)^{6} \,.  \eea By comparing the
coefficient of $z^{k+2}$ in Equation (\ref{charpoly}) we find a
necessary condition for the intersection numbers $\mu_{ab}$: the
trace of $H$ should be equal to the Euler characteristic of the
del Pezzo surface, \bea \mbox{Tr}H = k+3\,. \label{cons1}\eea This
gives us a Diophantine equation satisfied by the soliton spectrum
of the del Pezzo surfaces. This equation will turn out to play an
important role in the study of Seiberg dualities for the given
singularity.

We can use this formalism to calculate charges of fractional
branes. This is demonstrated in the following two examples:

 {\bf
Example one:} Consider the case of $\mathbb{B}_{4}$
($\mathbb{P}^{2}$ blownup at four points). The basis of
$H_{2}(\mathbb{B}_{k})$ we will consider is given by
$\{H,E_{1},\cdots , E_{k}\}$ with \bea H\cdot H=1\,,\,E_{a}\cdot
E_{b}=-\delta_{ab}\,,\,\,H\cdot E_{a}=0\,. \eea
The following collection of bundles and sheaves is an exceptional collection
forming a helix on $\mathbb{B}_{4}$ \cite{HIV, HI}
\bea &&\{-{\cal
O}(-H+E_{1}+E_{2}),{\cal O}_{E_{2}}(0),{\cal O}_{E_{1}}(0),{\cal
O}(E_{3}),{\cal O}(-H+E_{3}),\\ \nn &&\qquad -{\cal
O}_{E_{4}}(-1),-{\cal O}(-E_{4})\} \eea

Following \cite{HIV} we define
\bea
S_{ab}=\int_{\mathbb{B}_{4}}\mbox{ch}(F_{a}\otimes
F_{b}^{*})\mbox{Td}(\mathbb{B}_{4})\,.  \eea

It is easy to see that the characteristic polynomial of
$H=S^{-T}S$ is given by \bea
P(z):=\mbox{det}(z-H)=(z-1)^{7}=(z-1)^{\chi(\mathbb{B}_{4})}\,.
\eea Thus the Ramond charges are integers and are given
by \cite{CV},\footnote{The integer part can be calculated as shown in
\cite{CV}.} \bea
q_{R}=\{-1,0,0,0,0,0,1\}=\{\frac{p+q-2}{2}~|~p+q=\mbox{deg}\,\omega_{i},\,\omega_{i}\in H^{p,q}
(\mathbb{B}_{4})\}\,.  \eea
Where $\omega_{i}$ form a basis of $H^{p,q}(\mathbb{B}_{4})$.

{\bf Example two:} As another example we consider the case of
$\mathbb{B}_{8}$. In this case we consider the following
exceptional collection \cite{HI}, \bea\nn &&\{{\cal
O}(-H+E_{1}+E_{2}+E_{7}),{\cal O}_{E_{7}},{\cal O}_{E_{2}},{\cal
O}_{E_{1}},{\cal O}_{H-E_{3}-E_{6}},{\cal O}(-H+E_{6}),\\ \nn
&&\qquad{\cal O}(-E_{5}),-{\cal O}_{H-E_{3}-E_{5}}(1),-{\cal
O}_{E_{4}}(-1),-{\cal O}_{E_{8}}(-1),-{\cal O}(-E_{4}-E_{8})\}\,.
\eea
The characteristic polynomial of $H=S^{-T}S$, where $S$ is defined
as before, is \bea
P(z)&=&(z-1)^{11}=(z-1)^{\chi(\mathbb{B}_{8})}\,. \eea And the
Ramond charges are given by
($p+q=\mbox{deg}H^{p,q}(\mathbb{B}_{8})$) \bea
q_{R}&=&\{-1,0,0,0,0,0,0,0,0,0,1\}=\{\frac{p+q-2}{2}\,|\,p+q=\mbox{deg}
\,\omega_{i},\,\omega_{i}\in H^{p,q}(\mathbb{B}_{8})\}\,. \eea
Where $\omega_{i}$ form a basis of $H^{p,q}(\mathbb{B}_{8})$.

%
%
As discussed in detail in \cite{HIV} these exceptional collections
are not unique and other exceptional collections can be obtained
by mutations. However, all exceptional collections must give same
Ramond charges and therefore Equation (\ref{cons1}) must be
satisfied. This gives a severe constraint on the integers
$S_{ab}$. In the case of $\mathbb{B}_{0}:=\mathbb{P}^{2}$, which
is the compact divisor of the resolution of the Abelian orbifold
$\IC^3/\IZ_3$, it is easy to see that the equation is given by
\cite{CV} \bea \label{B0}
\mu_{21}^{2}+\mu_{31}^{2}+\mu_{32}^{2}+\mu_{21}\mu_{31}\mu_{32}=0\,,
\eea upon which we shall elaborate in Section \ref{glosup}.

\section{$\mathbb{C}^{3}/\mathbb{Z}_{N}$}
\label{Znorb}
 The theories arising on the D3-brane transverse to
an orbifold can be studied using the orbifold methods
\cite{orbifold}. In this section, however, we will use mirror
symmetry to study these theories and their Seiberg duals following
\cite{seiberg,CFIKV} where the case of
$\mathbb{C}^{3}/\mathbb{Z}_{3}$ was discussed.
\subsection{Weighted projective spaces}
The singularity $\mathbb{C}^{3}/\mathbb{Z}_{N}$ is produced by a
collapsing two complex dimensional weighted projective space. This can
be seen by using the linear sigma model description of the Calabi-Yau
threefold \cite{phases}. The linear sigma model charges of the
$\mathbb{C}^{3}/\mathbb{Z}_{N}$ are \cite{tachyon} \bea
(-N,k_{1},k_{2},k_{3})\,,\,\,k_{i}>0\,,\,\,k_{1}+k_{2}+k_{3}=N\,.
\eea The compact divisor described by the charges
$(k_{1},k_{2},k_{3})$ is the weighted projective space
$\mathbb{P}^{2}_{[k_{1},k_{2},k_{3}]}$, with homogeneous coordinates
\bea
[w_{1},w_{2},w_{3}]=[\lambda^{k_{1}}w_{1},\lambda^{k_{2}}w_{2},
\lambda^{k_{3}}w_{3}]\in
\mathbb{P}^{2}_{[k_{1},k_{2},k_{3}]}\,,\,\,\lambda\in
\mathbb{C}^{\times}\,.
\eea

We will consider the case when one of the $k_{i}$ is equal to one,
$k_{i}=(1,a,b)$. The weights $(1,a,b)$ give the action of
$\mathbb{Z}_{N}$ on the complex coordinates $z_{1},z_{2},z_{3}$ of
$\mathbb{C}^{3}$, \bea (z_{1},z_{2},z_{3})\mapsto (\omega
z_{1},\omega^{a}z_{2},\omega^{b}z_{3})\,,\,1+a+b=N\,,\,\omega=e^{\frac{2\pi
i}{N}}\,. \eea The corresponding weighted projective space is
$\mathbb{P}^{2}_{[1,a,b]}$. This weighted projective space is a
toric variety with the toric diagram in Figure (\ref{f:p^1ab}).
\onefigure{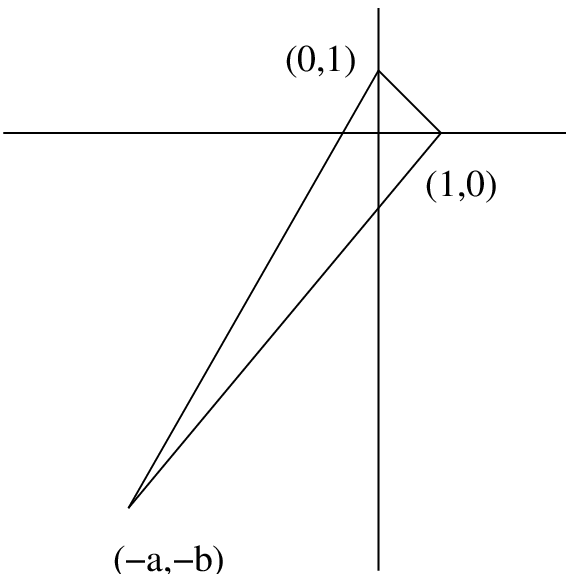}{Toric
diagram of $\mathbb{P}^{2}_{[1,a,b]}$.
\label{f:p^1ab}
}
We denote by $D_{1}, D_{2}$
and $D_{3}$ the divisors corresponding to the three faces. These
divisors are not all independent and satisfy the following relations
\bea D_{1}=aD_{3}\,,\,\,\,D_{2}=bD_{3}\,.  \eea
The intersection
numbers, which are useful when dealing with fractional branes, are
given by (defining $H=abD_{3}$)
\bea D_{1}\cdot
D_{1}&=&\frac{a}{b}\,,\,D_{1}\cdot D_{2}=1\,,\,D_{1}\cdot H=a\,,\\\nn
D_{2}\cdot D_{2}&=&\frac{b}{a}\,,\,D_{2}\cdot H=b\,,\,H\cdot H=ab\,,
\eea
and the cycle dual to the first Chern class is given by
\bea
c_{1}=D_{1}+D_{2}+D_{3}\,.
\eea
The $(p,q)$ web description of the above orbifolds is easy to obtain.
We consider the case for odd $N=2k+1$ for simplicity.
In this case we are looking
for a web with three external legs such that the intersection number of the
charges of the external legs is equal to $N$. If we let the external
charges\footnote{$(p_{1},q_{1})$ has been fixed using $\sl2z$.},
be $(p_{1},q_{1})=(-1,-1), (p_{2},q_{2})$ and $(p_{3},q_{3})$
then
\bea
\mbox{det}\begin{pmatrix}p_{2}& -1\\ q_{2}& -1\end{pmatrix}=
\mbox{det}\begin{pmatrix}-1 & p_{3}\\
-1&
q_{3}\end{pmatrix}=\mbox{det}\begin{pmatrix}p_{3}&p_{2}\\q_{3}&q_{2}
\end{pmatrix}=2k+1\,.
\eea The solution is given by\footnote{Up to $\sl2z$
transformations fixing $(-1,-1)$.}  \bea
(p_{2},q_{2})=(-k,k+1)\,,\,\,(p_{3},q_{3})=(k+1,-k)\,,\,\,
\sum_{i}p_{i}=\sum_{i}q_{i}=0\,.  \eea The resolution of the
singularity corresponds to resolving the web diagram as shown in
Figure (\ref{webz5}) and there could be many possible ways of
doing so corresponding the possible ways of orbifold action of
$\mathbb{C}^{3}$. In the case of $\mathbb{C}^{3}/\mathbb{Z}_{5}$
the resolution is unique and is determined completely by the
charges of the external legs of the web diagram.
\onefigure{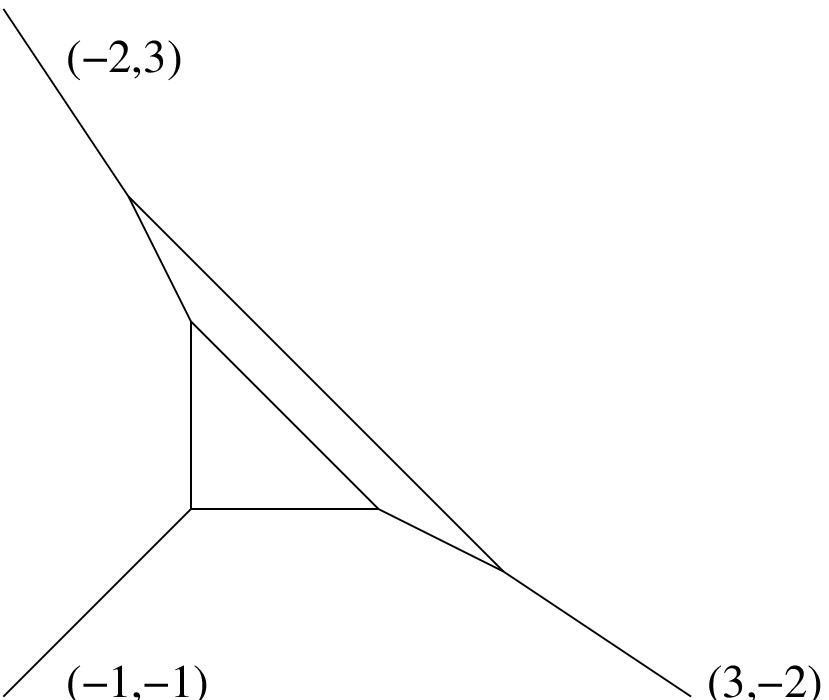}{The web diagram of the resolution of
$\mathbb{C}^{3}/\mathbb{Z}_{5}$.}
\subsection{Mirror manifold and fractional branes}
 From the linear sigma model description we can determine the
mirror Calabi-Yau as discussed in detail in \cite{HV,HIV}. In the
case we are interested in the linear sigma model charges are given
by $(-1-a-b,1,a,b)$ and the superpotential of the mirror LG theory
is \cite{HV} \bea
W&=&x_{0}+x_{1}+x_{2}+e^{-t}\frac{x_{0}^{1+a+b}}{x_{1}^{a}x_{2}^{b}}
\,,\,\,x_{i}=e^{-Y_{i}}\,, \eea where $t$ is the complexified
K\"ahler parameter, which measures the size of projective weight
space $\mathbb{P}^{2}_{[1,a,b]}$,
 and $Y_{i}$ are the fundamental fields taking
value in $\mathbb{C}$.  From the above superpotential the
following equation for the mirror Calabi-Yau can be determined
(taking $u,v,x,y \in \mathbb{C}$) \cite{HV,HIV}: \bea
e^{u}+e^{v}+e^{-au-bv}&=&z\,, \label{massive}\\ \nn
z+e^{\frac{t}{a+b+1}}&=&xy\,. \eea By homogenizing the first
equation we see that it defines a genus $g=\frac{(a+b)(a+b-1)}{2}$
curve over the z-plane. In the first equation the left hand side
of the equation is exactly the Landau-Ginsburg superpotential for
the massive ${\cal N}=(2,2)$ theory which is mirror of the sigma
model with weighted projective space $\mathbb{P}^{2}_{[1,a,b]}$ as
the target. Therefore from the discussion of Section \ref{N2} it
follows that the number of 3-cycles in the mirror geometry
described by Equation (\ref{massive}) is exactly equal to the
number of vacua of the massive LG theory and also the soliton
number between the vacua  gives the intersection number of
3-cycles\footnote{For a detailed discussion of how the 3-cycles
are
    constructed
in this geometry see \cite{HIV,HaIq,CFIKV}.} .Thus the quiver
diagram is given by the soliton diagram of the massive theory.

The soliton
diagram can be obtained easily for these geometries using the results of
\cite{HIV} and has been worked out in \cite{AV}.
\onefigure{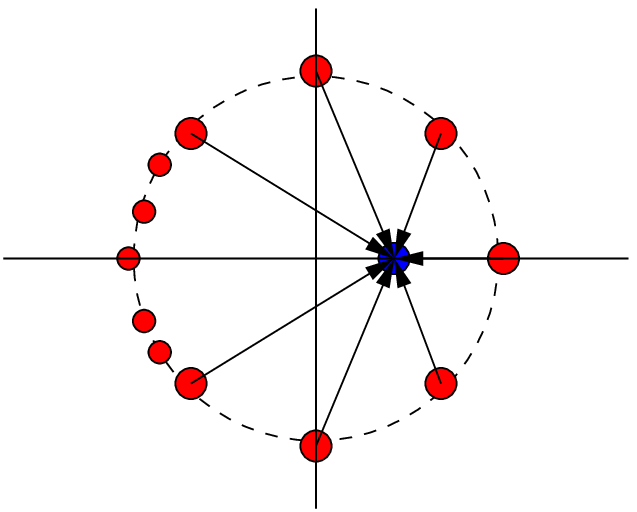}
{
\label{f:weighted}
Three cycles mirror to the fractional branes in
the resolution of $\mathbb{C}^{3}/\mathbb{Z}_{N}$.}

The vector bundles in this geometry mirror to the $S^{3}$'s are shown in
\figref{weighted} and are given by \cite{mayr,mayr2,OFS,AV}:
\bea
F_{i}=(S^{-1})_{ji}{\cal O}(j)\,,
\eea
where \cite{AV}
\bea
S_{ij}=\chi({\cal O}(i),{\cal O}(j))&=&0\,,\,\,\mbox{if}\,\,i<j\,\\ \nn
&=& \#\{(m_{1},m_{2},m_{3})\,|\,m_{i}\geq 0,i-j=m_{1}+am_{2}+bm_{3}\}.
\eea
Then we see that \bea
\chi(F_{i},{\cal O}(j))&=&\delta_{ij}\,,\\\nn
\chi(F_{i},F_{j})&=&(S^{-1})_{ij}\,,  \eea

For a weighted projective space
with weights $\{a_{1},a_{2},\cdots,a_{n}\}$ the generating function of
the number of solitons is given by \cite{AV} as
\bea F(q)=1+\sum_{I}
n_{I}q^{I}=\prod_{i=1}^{n}(1-q^{a_{i}})\,.
\eea
The number of
solitons between the $i$-th and the $j$-th vacua can be read from the above
function as $n_{|i-j|}$. For the case we are interested in
\bea
\label{Fq}
F(q)&=&\prod_{k=1}^{3}(1-q^{a_{k}})=(1-q)(1-q^{a})(1-q^{b}) \,,\\ \nn
&=&1-q-q^{a}+q^{a+1}-q^{b}+q^{b+1}+q^{a+b}-q^{a+b+1}\,.
\eea
And this
gives us the  quiver diagram for each $i$ in Figure (\ref{quiver}).
Note that
 in the case $b=a+1$ (i.e., $N=2a+2$) there are  bi-directional arrows
related to nodes.
\onefigure{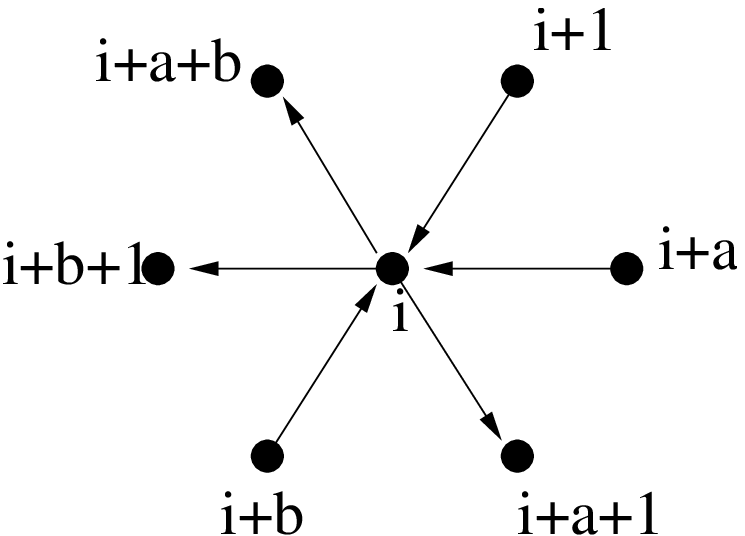}{The $i$-th node of the quiver diagram for
$\IC^3/\IZ_N$. We have marked all the nodes linked to $i$.}
There are 8 terms in Equation (\ref{Fq}) where the first term and
the last term do not contribute to the quiver diagram and cancel
each other since we identify the nodes modulo $N$. Now $F(1)=0$
implies the anomaly cancellation condition (the number of incoming
and outgoing arrows being the same for each node).
In general the mirror manifold is given by a (non-compact) genus $g$
fibration over the z-plane and a $\mathbb{C}^{\times}$ fibration over
the z-plane. The genus $g$ fibration degenerates at $N=\chi(M)$ number
of points on the z-plane where $M$ is compact divisor of $X$. In
general $M$ will be a set of four manifolds $M_{i}$ joined together along some
rational curves. The degeneration, of the genus $g$ fibration, is due
to a 1-cycle collapsing. Using these collapsing 1-cycles one can
construct 3-cycles, which are topologically $S^{3}$, in the mirror
manifold $Y$.

Let us denote by $V_{a}$ the set of exceptional bundles on $M$
corresponding to the fractional branes, \bea\nn
\mbox{ch}(V_{a})=\{(r^{(a)}_{1},\cdots,r^{(a)}_{g}),(\Sigma^{(a)}_{1},\cdots,
\Sigma^{(a)}_{g}),(k^{(a)}_{1},\cdots
k^{(a)}_{g})\}\,,\,\,a=1,\cdots,N.  \eea Where $r_{i}^{(a)}$ is
the rank, $\Sigma^{(a)}_{i}$ the first Chern class and
$k^{(a)}_{i}$ the second Chern character of the restriction of the
bundle $V^{(a)}$ to $M_{i}$. Given this set of fractional branes
the set of vanishing 1-cycles ${\cal C}_{a}$ is given
by\footnote{Here $\{\alpha_{1},\beta_{1},\cdots,
\alpha_{g},\beta_{g}\}$ is a basis of 1-cycles on the genus $g$
curve such that the only non-zero intersection numbers are
$\alpha_{i}\cdot \beta_{i}=1,\,i=1,\cdots,g$.}  \bea {\cal
C}_{a}=\sum_{i=1}^{g}d_{\Sigma^{(a)}_{i}}\alpha_{i}+\sum_{i=1}^{g}
r^{(a)}_{i}\beta_{j}\,\,,\,a=1,\cdots,N\,. \label{charges} \eea It
follows then that the quiver diagram given by the intersection
matrix of 3-cycles is
 \bea I_{ab}={\cal C}_{a}\cdot {\cal
C}_{b}=\sum_{i=1}^{g}\mbox{det}\begin{pmatrix}d^{(a)}_{i} &
d^{(b)}_{i}\\ r^{(a)}_{i}& r^{(b)}_{i}\end{pmatrix}\,.  \eea
And if
the sum of the fractional branes is a D3-brane (0-cycle),
\bea
\sum_{a=1}^{N}\,r^{(a)}_{i}=\sum_{a=1}^{N}d_{\Sigma^{(a)}_{i}}=0\,,\,\forall
i=1,\cdots,g\,.
\eea
then ${\cal C}=\sum_{a=1}^{N}{\cal
C}_{a}$ is such that
\bea
{\cal C}\cdot {\cal C}_{a}=0\,,\,\forall
a=1,\cdots,N\,,
\eea
and gives rise to a $T^{3}$.

\subsection*{Example: $\mathbb{C}^{3}/\mathbb{Z}_{5}$:} We consider the
case when $(a,b)=(2,2)$ (the other case of $(a,b)=(1,3)$ is equivalent
to this one). In this case we have $\mathbb{P}^{2}_{[1,2,2]}$ as the
weighted projective space collapsing to produce the orbifold
singularity.  The weighetd projective space itself has singularities
which when resolved give the compact divisor $M$ as a $\mathbb{P}^{2}$
and $\mathbb{F}_{2}$ joined along a rational curve \cite{OFS} as can
be seen in the web diagram Figure (\ref{webz5}). As discussed in the
previous section the mirror Calabi-Yau is a genus two fibration and a
$\mathbb{C}^{\times}$ fibration over the z-plane given by
\bea
W:=e^{u}+e^{v}+e^{-2u-2v}+e^{-t_{2}+\frac{3}{5}t_{1}-v}&=&z\,,\\ \nn
z+e^{t_{1}/5}&=&xy\,.
 \eea
In the above equation $t_{1}$ and $t_{2}$
are the complexfied K\"ahler parameters. The genus two fibration
degenerates at five points on the z-plane (depending on the K\"ahler
parameters) as can be seen by solving the equations
$\partial_{u}W=\partial_{v}W=0$. For
$\lambda_{2}=e^{-t_{2}+(2/5)t_{1}}<<1$ the five degenerate fibers lie
approximately on a circle in the z-plane and
are, using Equation (\ref{charges}) and results of \cite{OFS} \footnote{Here
$\{\alpha_{1},\beta_{1},\alpha_{2},\beta_{2}\}$ is a basis of 1-cycles
on the genus two curve such that the only non-zero intersection
numbers are $\alpha_{1}\cdot \beta_{1}=\alpha_{2}\cdot \beta_{2}=1$.}
\bea C_{1}&=&-\beta_{1}+2\beta_{2}\,, \,\,
C_{2}=-\alpha_{1}-3\beta_{2}\,,\,\,
C_{3}=4\beta_{1}+\alpha_{2}-\beta_{2}\,,\\\nn
C_{4}&=&-\beta_{1}-\alpha_{2}\,,\,\,
C_{5}=\alpha_{1}-2\beta_{1}+2\beta_{2}
\,. \eea
Given these cycles the
quiver diagram can be obtained from the intersection numbers (see Figure
(\ref{quiver-z5})),
\bea
S_{a}\cdot S_{b}=C_{a}\cdot C_{b}\,. \eea
\onefigure{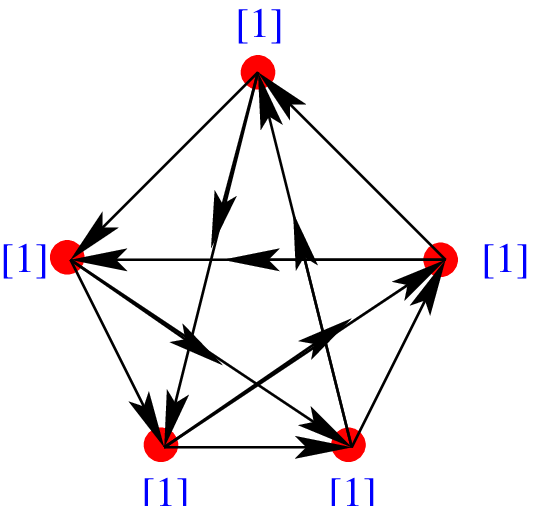}{The quiver diagram of
$\mathbb{C}^{2}/\mathbb{Z}_{5}$. Here we use $[1]$ to denote the rank of
that node is 1.}


\section{Local del Pezzo Surfaces}
\label{dPs}
\subsection{Mirror Manifolds and Elliptic fibration}
In this section we will consider non-compact Calabi-Yau threefolds
which are line bundles over del Pezzo surfaces and their mirror
partners. We will study the geometry of the mirror manifold in
detail and see that results about classification of $[p,q]$
7-branes in F-theory backgrounds actually allow us to write
Diophantine equations, for all del Pezzo surfaces, whose solutions
determine the quiver diagrams.  We will see that these Diophantine
equations derived from the geometry are the same as the equations
given by the theory of solitons in ${\cal N}=(2,2)$ massive
theories in two dimensions \cite{CV}.

The superpotential of the LG theory mirror to the linear sigma model
provides the description of mirror CY. Since $\mathbb{P}^{2}$ blown up
at more than three points is not toric we cannot use the linear sigma
model to obtain the mirror CY. However, it is possible to obtain the
mirror manifolds $X_{k}$ to local non-toric del Pezzos $Y_{k}$ along the lines discussed in \cite{HI} using the fact that
$\frac{1}{2}$-K3 ($\mathbb{P}^{2}$ blown up at nine points) is self-mirror
\cite{estring}. The Calabi-Yau manifold mirror to local $\mathbb{B}_{k}$
($\mathbb{P}^{2}$ blown up at $k$ points) is given by ($x,y,z,w_{1},w_{2}\in \mathbb{C}$)\bea
y^{2}&=&x^{3}+f_{k}(z)x+g_{k}(z)\,,\nonumber\\  w_{1}w_{2}&=&z-z_{*}\,,
\label{mirror}
\eea
where $f_{k}(z)$ and $g_{k}(z)$ are polynomials in $z$ and the
explicit form of these polynomials can be found in \cite{SZ}.
The parameters in the polynomials $f_{k}$ and $g_{k}$ are the complex
structure parameters of the mirror CY and are related to the K\"ahler
structure parameters of the local del Pezzo.

The geometry of the mirror manifold $X_{k}$ is clear and is
discussed in several papers \cite{HaIq, HIV, CFIKV}. We briefly
mention it here again for completeness and because it will be
useful for later discussion.

The first equation in (\ref{mirror}) describes an elliptic
fibration over the complex z-plane. This elliptic fibration has $k+3$
degenerate fibers whose positions depend on the K\"ahler parameters of
the Calabi-Yau $Y_{k}$ or the complex structure parameters of
$X_{k}$.
The second equation in (\ref{mirror}) describes a $\mathbb{C}^{*}$
fibration over the z-plane such that at $z=z_{*}$ the $\mathbb{C}^{*}$
fibration degenerates when
its non-trivial $S^{1}$ shrinks.

The only non-trivial compact closed
cycles in this geometry are 3-cycles. These 3-cycles are constructed
as follows: we connect the point $z_{*}$ to the position of the
degenerate fiber $z_{a}$, over this path we have 2-cycles which
collapse at the two ends of this interval.
The circle of the $\mathbb{C}^{*}$
fibration collapses at $z_{*}$ and a 1-cycle of the elliptic fibration
collapses at $z_{a}$. These cycles together with the path in the
z-plane form a closed 3-cycle which is topologically an $S^{3}$.

In this way we obtain $k+3$ 3-cycles with topology of $S^{3}$. This
lattice of
$k+3$ 3-cycles is mirror to $k+3$ dimensional lattice $H_{0}(\mathbb{B}_{k})\oplus H_{2}(\mathbb{B}_{k})\oplus H_{4}(\mathbb{B}_{k})$ of
compact cycles in ${\cal B}_{k}$. The intersection between the
3-cycles is
completely determined by the vanishing cycles of the elliptic
fibration and the point $z_{*}$ can be thought of as the point from
which the charges of the vanishing cycles are to be measured.
This is
quite reasonable since the only points at which the 3-cycle intersect
lie on the elliptic fiber above the point $z=z_{*}$. Thus if the
vanishing cycles are $C_{a}\equiv [p_{a},q_{a}]\in
H_{1}(\pi^{-1}(z_{*}),\bbbz)$ and the corresponding 3-cycles are
$S_{a}$ then \bea S_{a}\cdot S_{b}:=C_{a}\cdot C_{b}=\mbox{det}
\begin{pmatrix} p_{a} & p_{b} \\ q_{a} & q_{b}\,\end{pmatrix}.  \eea
Thus the information about the intersection numbers is naturally
contained in the charges of the vanishing cycles of the elliptic
fibration. Because of Picard-Lefschetz monodromy there is no
unique choice of charges and by changing the paths connecting the
position of degenerate fibers to $z_{*}$ we can change the
charges.

A natural question is whether there is some invariant which
characterizes this configuration of degenerate fibers. As
discussed at length in \cite{papers} the only invariants of these
configurations are the number of the degenerate fibers, the trace
of the $SL(2,\mathbb{Z})$ monodromy matrix and the greatest common
divisor of the intersection numbers. Actually one can write down
Diophantine equations such that their solutions completely
describe all the configurations which can be obtained by
Picard-Lefschetz transformations as was done in \cite{CFIKV} for
the case of $Y_{0}$.

It is easy to understand the origin of such an equation. The
monodromy matrix of a configuration of degenerate matrix is
invariant under Picard-Lefschetz transformations but it is not
invariant under global $SL(2,\mathbb{Z})$ transformation. However,
the trace of the monodromy matrix is invariant under global
$SL(2,\mathbb{Z})$ as well as Picard-Lefschetz transformations.
The trace of the monodromy matrix does not depend on the vanishing
charges and only depends on the intersection numbers between the
vanishing charges \cite{papers}. The configuration of degenerate
fibers we are considering are such that they have trace of the
monodromy matrix equal to two. The implications of this were
discussed at length in \cite{papers}. Thus the solutions of the
equation \bea \mbox{det}(K_{\mathbb{B}_{k}}-1)=\mbox{Tr}
K_{\mathbb{B}_{k}}-2=0\,, \eea except the trivial solution,
completely describe different configurations related by
Picard-Lefschetz transformations.

\subsubsection*{local $\mathbb{B}_{0}$:}
This case was discussed in
\cite{CFIKV} where the Diophantine equation was also given but was
derived using the relation of the local $\mathbb{B}_{0}$ mirror geometry with
superpotential geometry of the mirror of the massive $\mathbb{P}^{2}$
model. We will show that these two points of view give the same
equation in all local del Pezzo cases. The equation in this case is
\bea
\mu_{21}^{2}+\mu_{32}^{2}+\mu_{31}^{2}+\mu_{21}\mu_{32}\mu_{31}=0\,.
\eea

\subsubsection*{local $\mathbb{B}_{1}$ and local $\mathbb{F}_{0}$:}
In both these cases the equation is given by
\bea
\mu_{21}^{2}+\mu_{31}^{2}+\mu_{41}^{2}+\mu_{32}^{2}+\mu_{42}^{2}
+\mu_{43}^{2}+\mu_{21}\mu_{32}\mu_{31} \qquad \qquad \\ \nn
+\mu_{21}\mu_{42}\mu_{41}+\mu_{31}\mu_{43}\mu_{41} +\mu_{32}\mu_{43}\mu_{42}
+\mu_{21}\mu_{32}\mu_{43}\mu_{41}=0\,.  \eea
In order to distinguish the two
models, $\mathbb{B}_{1}$ and $\mathbb{F}_{0}$, we divide the solutions
of the above equation in two sets. One set will have solutions with
gcd equal to one giving the result for $\mathbb{B}_{1}$ and the other set has
solutions with gcd equal to two giving the result for the $\mathbb{F}_{0}$
case. At this moment we do not know why $\mathbb{B}_{1}$ and $\mathbb{F}_{0}$
are distinguished by gcd. However, recalling the fact that  $\mathbb{B}_{1}$
has only $SU(2)$ global flavor symmetry while $\mathbb{F}_{0}$ has
$SU(2)\times SU(2)$ symmetry \cite{multi}, we speculate that the symmetry
maybe the reason behind. Similarly, we speculate that the fact that
 the gcd of  $\mathbb{B}_{0}$ in all phases are always 3 is related to the
$SU(3)$ global flavor symmetry.

\subsubsection*{local $\mathbb{B}_{k}$:}
In the general case the equation is given by \cite{papers}\bea
TrK-2=\sum_{r=2}^{k+3}\sum_{i_{1}>i_{2}>i_{3}\cdots >i_{r}}
 \mu_{i_{1}i_{2}} \mu_{i_{2}i_{3}}\cdots
\mu_{i_{r-1}i_{r}}\mu_{i_{1}i_{r}}&=&0\,. \eea
Where $K$ is the monodromy matrix of a configuration of $k+3$ degenerate fibers.

>From the above equation it is clear that if $\mu_{ij}=0$ for all $j$
and a fixed $i$ then the equation reduces to the Diophantine equation
for $\mathbb{B}_{k-1}$. All solutions of the above equation are
Picard-Lefshetz equivalent to the intersection numbers obtained from
the following configuration, \bea \mu_{ij}&=&C_{i}\cdot C_{j}\,,\\ \nn
C_{i}&=&\alpha\,,\,i=1,\cdots,k\,,\,\,C_{k+1}=2\alpha-\beta\,,\,C_{k+2}=-\alpha+2\beta\,,\,\,C_{k+3}=-\alpha-\beta\,.
\eea

\subsubsection*{$\mathbb{C}^{3}/\mathbb{Z}_{5}$:}
For $\mathbb{C}^{3}/\mathbb{Z}_{5}$ we can also write down a
Diophantine equation whose solutions (except the trivial one) give
the quiver diagram and the gauge group factors for all theories
obtained from this geometry by Picard-Lefschetz transformation. If
$M$ is the monodromy matrix around five degenerate fibers of a
genus two fibration (given in terms of the charges of the
degenerate fibers) then the Diophantine equation, which is
function only of the intersection numbers, is given by \bea
\mbox{det}(M-\unit)=0\,.\label{Z5eq} \eea The above equation
simply means that the collection of degenerate fibers allow an
eigenvector of eigenvalue one. Thus if the above equation is
satisfied then there is a 1-cycle in the fibration which is
invarinat under the monodromy and gives topologically a $T^{2}$
together with the path in the base that goes around the degenerate
fibers. This $T^{2}$ together with the $S^{1}$ of the
$\mathbb{C}^{\times}$ fibration gives a $T^{3}$. In the case of
$\mathbb{C}^{3}/\mathbb{Z}_{3}$ and other local del Pezzo
singularities the matrix $M$ is an $\sl2z$ matrix and therefore
since for an $\sl2z$ matrix $\mbox{det}(M-\unit)=2-\mbox{Tr}M$,
the equation is given by $\mbox{Tr}M=2$ as discussed in the
previous section.

It is easy to show that two solutions to Equation (\ref{Z5eq}), which
are related to each other by PL transformation, are given by \bea
C_{1}&=&-\beta_{1}+2\beta_{2}\,, \,\,
C_{2}=-\alpha_{1}-3\beta_{2}\,,\,\,
C_{3}=4\beta_{1}+\alpha_{2}-\beta_{2}\,,\\\nn
C_{4}&=&-\beta_{1}-\alpha_{1}\,,\,\,
C_{5}=\alpha_{1}-2\beta_{1}+2\beta_{2} \,. \eea and \bea
\widehat{C}_{1}=-C_{1}\,,\,\,\widehat{C}_{2}=C_{2}\,,\,\,
\widehat{C}_{3}=C_{3}\,,\,\,\widehat{C}_{4}=C_{4}+2C_{1}\,,\,\,
\widehat{C}_{5}=C_{5}+C_{1}\,.
\eea On the z-plane the two solutions correspond to the 3-cycles shown
in Figure (\ref{ns}) below.
\onefigure{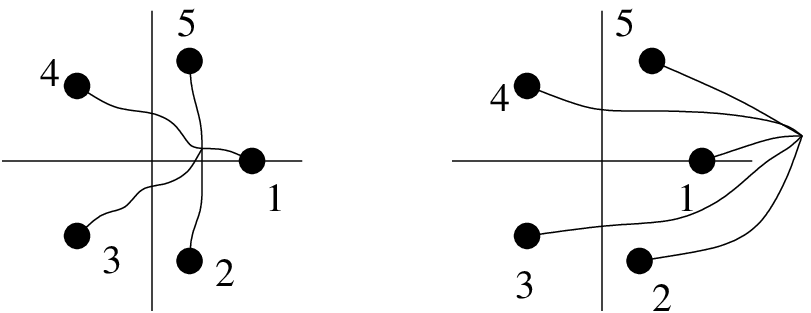}{Two solutions to the diphantine
equation related to each other by PL transformation.}
 These two
solutions are Seiberg dual to each other.  The first solution
gives a $\prod_{i=1}^{5}U(1)$ gauge theory with quiver given by
Figure (\ref{Z5})(a). The second solution gives a
$U(2)\times\prod_{i=1}^{4}U(1)$ gauge theory with quiver given by
Figure (\ref{Z5})(b). \onefigure{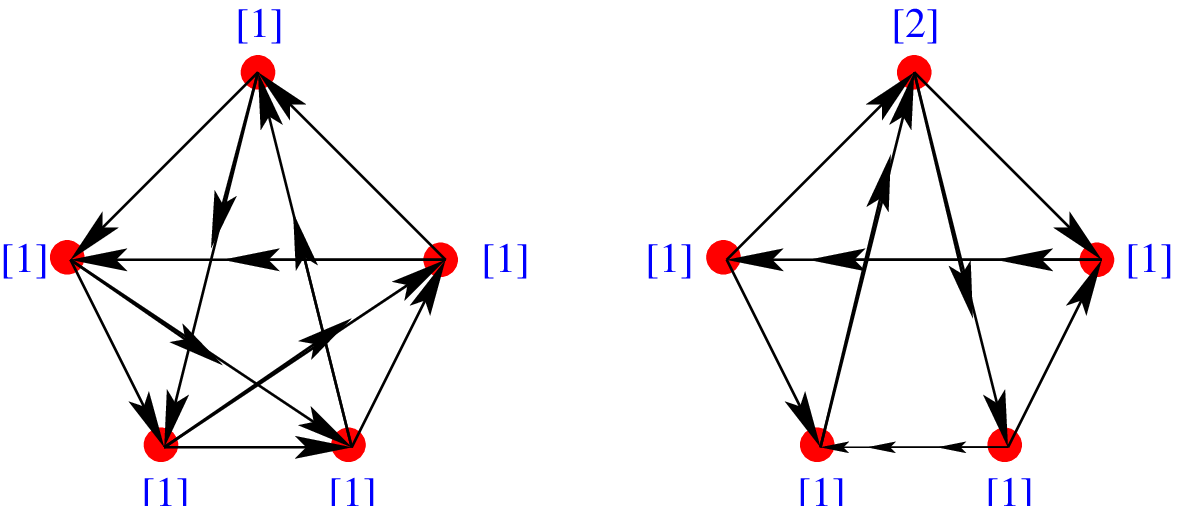}{Two Seiberg dual quivers
related to each other by Picard Lefschetz transformations.}

\subsection*{$\mathbb{C}^{3}/\mathbb{Z}_{2k+1}$:} In this case we
take the action of $\mathbb{Z}_{2k+1}$ on the $\mathbb{C}^{3}$ to be
given by $(1,k,k)$. In this case the compact divisor of the resolved
space is a $\mathbb{P}^{2}$ and $k-1$ Hirzebruch surfaces joined along
the rational curves. The mirror is given by \cite{HV,HIV} \bea
e^{u}+e^{v}+e^{-ku-kv}+\sum_{m=1}^{k-1}\lambda_{m}e^{-mu-mv}=z\,,\\
\nn z+e^{\frac{t_{0}}{2k+1}}=xy\,.  \eea Where $\lambda_{m}$ are
the complex structure parameters related to the K\"ahler
parameters as $\lambda_{m}=e^{-t_{m}+\frac{t_{0}}{2m+1}}$.  For
$|\lambda_{m}|<<1$ we see that the critical points of the first
equation above lie on a circle and the 3-cycles are as shown in
Figure (\ref{z2k})(a).  As we change the K\"ahler parameter
$t_{0}$ we see that cycles undergo Picard-Lefschetz transformation
as shown in Figure (\ref{z2k})(b). In this case also the
intersection matrix of cycles will obey the Equation (\ref{Z5eq})
with $M$ a $2k+1$ monodromy matrix given in terms of intersection
numbers. However, in this case since there are more than one
possible in equivalent resolutions of the singularity therefore
not all solutions will be related by PL transformations. Thus it
is necessary condition but not sufficient. \onefigure{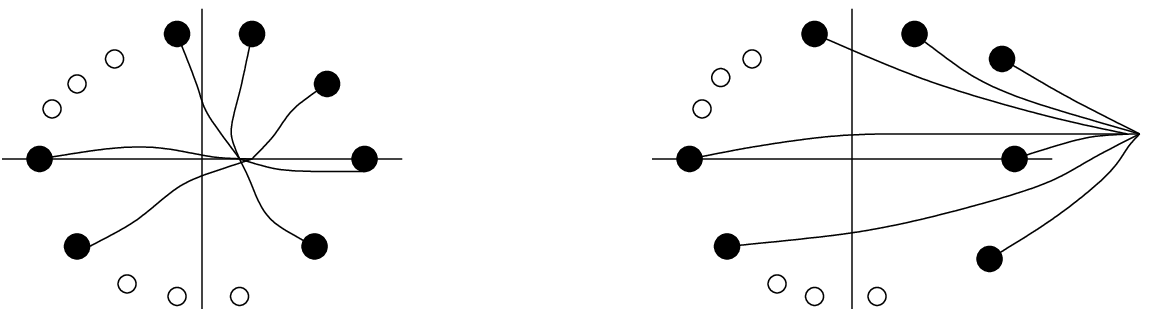}{The two
basis for the $\mathbb{C}^{3}/\mathbb{Z}_{2k+1}$.} Under the
change in the basis of cycles shown in Figure (\ref{z2k})(b) the
new basis is given by \bea S_{0}&\mapsto&
\widehat{S}_{0}=-S_{0}\,,\\ \nn S_{i}&\mapsto&
\widehat{S}_{i}=S_{i}\,,\,\,i\neq 0, k+1,2k\,,\\ \nn
S_{k+1}&\mapsto& \widehat{S}_{k+1}=S_{k+1}+2S_{0}\,,\\ \nn
S_{2k}&\mapsto& \widehat{S}_{2k}=S_{2k}+S_{0}\,.  \eea Where we
have used the intersection numbers between the 3-cycles determined
from the quiver diagram given by $F(q)=(1-q)(1-q^{k})^{2}$. Since
$\sum_{a=0}^{2k} S_{a} =[T^{3}]$ therefore requiring the same for
the new basis we get \bea\
\sum_{a=0}^{2k}n_{a}\widehat{S}_{a}=[T^{3}]\,\,\implies\,\,\,n_{0}=2,\,n_{i>0}=1\,.
\eea Thus we get  a $U(2)\times \prod_{i=1}^{2k}U(1)$ theory with
quiver determined from $\widehat{S}_{a}\cdot \widehat{S}_{b}$.  It
is easy to check that the new quiver is indeed that of the Seiberg
dual theory with duality performed on the $0-th$ node
corresponding to $S_{0}$.

\section{Seiberg Duality and Picard-Lefschetz Monodromy}
\label{five}
 The realisation that Seiberg's duality can be geometricised as
Picard-Lefschetz monodromy has been permeating in the literature since
at least \cite{OV2}. Recently works on Toric Duality
\cite{toric,toric2,seiberg,chris} have beckoned for a re-examination
of the geometry of Seiberg duality. Indeed some ideas were presented
in \cite{seiberg} and addressed in detail in \cite{HaIq,CFIKV}. The
purpose of this section is to explicit the derivation, as mentioned in
\cite{HaIq}, of Seiberg duality in terms of the quiver rules in
\cite{seiberg} and point out some interesting examples in a
comprehensive fashion. In due course we will resolve the discrepancies
and puzzles which arose in \cite{seiberg} concerning the relation
between Seiberg duality and Picard-Lefschetz theory. The relation
between Seiberg duality and Picard-Lefshetz transformation was
discussed in detail in \cite{CFIKV} in terms of mutation of bundles
and the corresponding action on the 3-cycles.

Instead of using the nomenclature of \cite{seiberg}, we shall here
use the language of $(p,q)$ 7-branes. As addressed in the earlier
sections, the mirror picture of the transverse D-brane probe on
$M$ consists of collections of (vanishing) three cycles $S_i$ with
$n_i$ D-branes wrapped thereon in the mirror Calabi-Yau $W$. The
result is a $\prod_i U(n_i)$ gauge theory with bifundamentals
$a_{ij}$ given as the intersection matrix of these vanishing
cycles $S_i \cdot S_j$.

Now let us phrase these (Picard-Lefschetz) cycles $S_i$ in the
language of $(p,q)$ 7-branes in the spirit of Section \ref{N2},
emphasizing on the monodromy. Each $S_i$ can be represented by a
$[p_i,q_i]$ 7-brane together with a wrapping number
$n_i$\footnote{Comparing with equation (\ref{t3}), here we use
$n_i$ instead of $n_a$ to emphasize that we consider the general
situation where the sum does not need to be $T^3$.}. The usual
anomaly cancellation condition $\sum_ja_{ij} n_j = 0$ now
translates to
\begin{equation}
\sum\limits_j (S_i \cdot S_j) n_j =0, \qquad \qquad \forall j.
\label{anomaly}
\end{equation}
In other words, the cycle $T=\sum\limits_j n_j S_j$ should have
zero intersection with any cycle $S_i$.
One particular case is that the cycle $T$ is precisely the $T^3$
fibre. As far as the $(p,q)$ charges are concerned,
(\ref{anomaly}) is simply (see Equation (\ref{AIJ})) \bea
\sum\limits_i n_i [p_i, q_i]=0. \label{npq} \eea

As mentioned earlier the bifundamental matter content is given by
intersection numbers, which are computed as determinants. Strictly
speaking, in the usual notation the adjacency matrix $a_{ij}$ of
the quiver is such that $A_{ij}=a_{ij} - a_{ji} = S_i \cdot S_j$
where to distinguish them we have used $A_{ij}$ which can be both
positive or negative. Through out the whole paper, except in
Section \ref{52}, we assume that only one of $a_{ij},a_{ji}$ of a
given pair $i,j$ is nonzero. Under this assumption, if $A_{ij}>0$
it is $a_{ij}\neq 0$ and if $A_{ij}<0$ it is $a_{ji}\neq 0$. The
$A_{ij}$ is calculated as
\begin{equation}
\label{AIJ}
 A_{ij} = S_i \cdot S_j=
\left|
\begin{array}{cc}
p_i & q_i \\ p_j &  q_j
\end{array}
\right|.
\end{equation}

Now Picard-Lefschetz monodromy is the motion of vanishing cycles $S_j$
about a chosen one $S_i$ such that thereafter the cycles becomes the
linear combination (no summation on $i$)
\[
S_j \rightarrow S_j + (S_j \cdot S_i) S_i.
\]
With wrapped branes the situation is a little more involved as we
have to take into account how fractional branes rearrange in the
new basis. Alternatively, one can take into account the usual
brane creation mechanism \cite{HIV,CFIKV}. However before we
elaborate on how to cooperate these factors in the following
discussion, let us restate Picard-Lefschetz transformations in the
language of $(p,q)$ charges:
\subsection*{Rules for Picard-Lefschetz on $(p,q)$ 7-branes}
\begin{enumerate}
\item For a collection of vanishing cycles $S_i = [p_i,q_i]$, each with
    wrapping number $n_i$, we let $S_{k-1}$ pass through
    $S_k$ for a chosen $k$.
\item All $S_i$ and $n_i$ for $i \ne k,k-1$ remain uneffected.
\item The concerned cycles transform as $S_{k-1} \rightarrow
S_{k-1} + (S_{k-1} \cdot S_k) S_k$ and $S_k \rightarrow S_k$,
i.e., \bea
    \begin{pmatrix}
    p_{k-1} \\ q_{k-1}\end{pmatrix} & \rightarrow &
    \begin{pmatrix}p_{k-1} \\ q_{k-1}\end{pmatrix}
    +[Det\begin{pmatrix}p_{k-1}& q_{k-1}\\ p_k & q_k \end{pmatrix}]
    \begin{pmatrix}p_{k} \\ q_{k}\end{pmatrix}\\
    &  = &
    \begin{pmatrix} 1 + p_{k} q_{k} & -p_{k}^2 \\ q_{k}^2 & 1 -
    p_{k} q_{k}\end{pmatrix}
    \begin{pmatrix} p_{k-1} \\ q_{k-1} \end{pmatrix};\\
    \begin{pmatrix} p_k \\ q_k\end{pmatrix} & \rightarrow &
    \begin{pmatrix} p_k \\ q_k\end{pmatrix}
    \eea

\item   The wrapping numbers transform as
    \[
    \begin{array}{cc}
    n_{k-1} \rightarrow n_{k-1};
    &
    n_k \rightarrow n_k - (S_{k-1} \cdot S_k) n_{k-1}
    \end{array}
    \]
    If this new number $n_k$ is negative,
    we should simply make it positive and multiply the new
    $[p_k,q_k]$ by $-1$. In geometric language, this changes
    the fractional brane into the anti-fractional brane as well as
    the
    orientation of the cycle $S_k$,\footnote{When we apply this to
    Seiberg duality, it is more convenient to define $S_k\rightarrow
    -S_k$ and $n_k\rightarrow -n_k+(S_{k-1} \cdot S_k) n_{k-1}$.}
    so the net effect of (anti)-branes wrapping will be the same.
\item Now we have a new collection $(\widetilde{n_i},\widetilde{S_i})$
    which is as before
    except for when $i = k,k-1$; these could then be used to
    calculate the new quiver $\widetilde{S_i} \cdot\widetilde{S_j}$.
\end{enumerate}

As a check let us verify that the anomaly cancellation still holds.
The condition (\ref{anomaly}) now reads
\beqan
\sum_i \widetilde{n_i} \widetilde{S_i}
    & = & \sum_{i\neq k,k-1} n_i S_i + n_{k-1} (S_{k-1} + (S_{k-1}
    \cdot S_k) S_k + (n_k - (S_{k-1} \cdot S_k) n_{k-1}) S_k \\
    & = & \sum_i  n_i S_i +  n_{k-1} S_k (S_{k-1} \cdot S_k) -
        (S_{k-1} \cdot S_k) n_{k-1}  S_k \\
    & = & \sum_i  n_i S_i \\
    & = & 0,
\eeqan which is as desired. The new quiver remains anomaly-free
after any Picard-Lefschetz transformation.
\subsection{Example: Hirzebruch Zero}
After these generalities let us re-examine the by now familiar example
of the cone over the zeroth Hirzebruch surface $\mathbb{F}_{0}$
\cite{toric,toric2,seiberg,CFIKV}. The two toric (Seiberg) dual cases
are recapitulated in Figure (\ref{F0}). Our starting point is the
following set of $(p,q)$ charges giving the affine $E_{1}$ background
\cite{papers,HI}
$$
A:~N+M_1~[1,-1] \qquad B:~N+M_2~[1,1] \qquad C:~N+M_1~[-1,1]
\qquad D:~N+M_2~[-1,-1]
$$
where $N,M_1,M_2$ are all positive integers. This is the most
general form of anomaly free theories on $\mathbb{F}_{0}$. Notice
also that although we have three numbers $N,M_1,M_2$, only two
combinations, say $|M_1-M_2|$ and $N+min(M_1,M_2)$, are
independent parameters.
 We can easily verify by computing pairwise
determinants that the quiver is as given in Case (I).

\onefigure{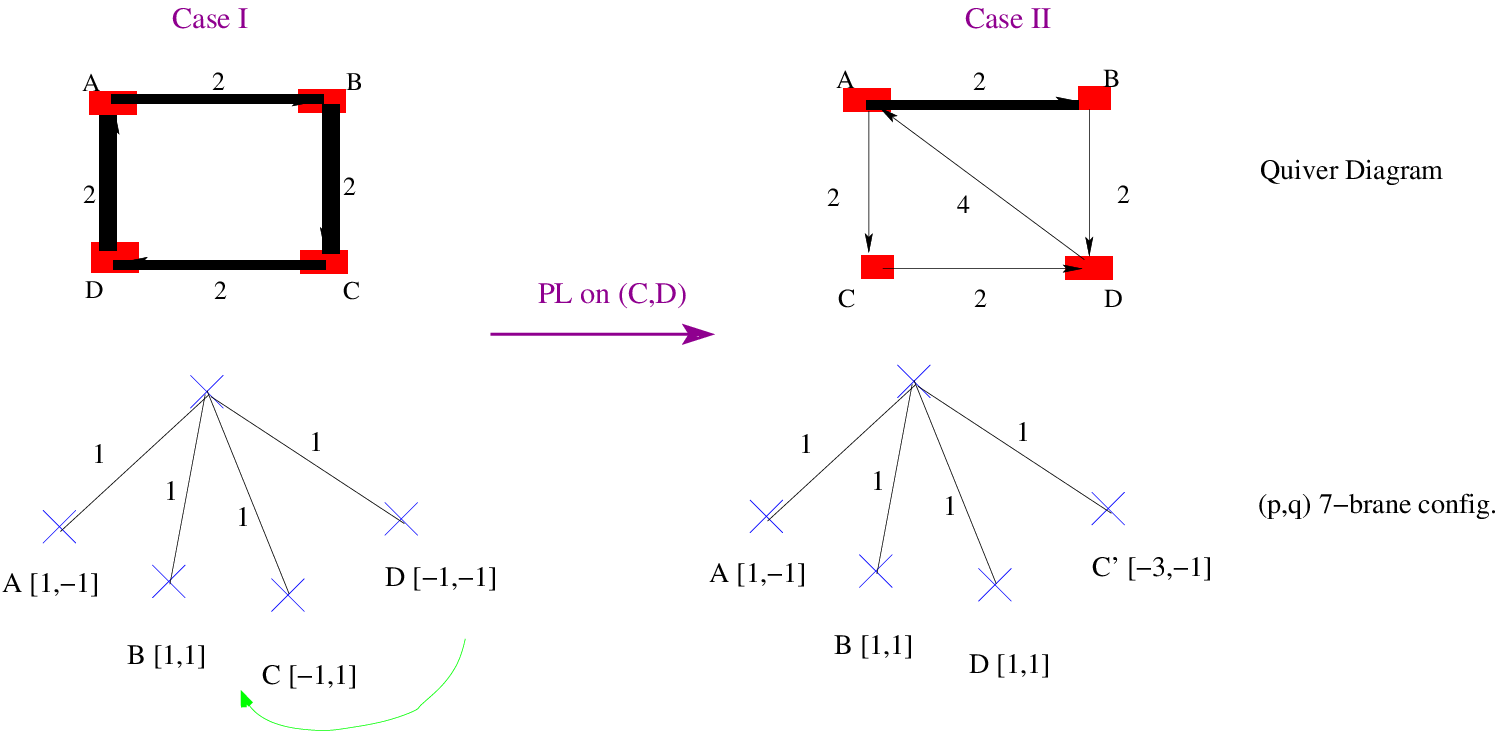}{
The toric (Seiberg) dual cases of the zeroth Hirzebruch
    surface $F_0$. Below the respective quiver diagrams we show
    the $(p,q)$ 7-brane configurations which give the
    quiver. Furthermore we see clearly how a Picard-Lefschetz
    transformation,
    as we move $D$ across $C$ in Case (I), gives Case (II).}

If we move cycle $D$ past $C$, we will obtain the new configuration
$$
A:~N+M_1~[1,-1]~~~~B:~N+M_2~[1,1]~~~~D:~n_D~[-1,-1]~~~~~C':~N+M_1~
$$
where according to the rules above $A,B,D$ remain unchanged while
$$
C \rightarrow C'=
\left[
\begin{array}{cc}
1+p q  &  -p^2 \\ q^2 & 1-pq
\end{array}
\right]_D \left[\begin{array}{c} p \\ q \end{array} \right]_C =
\left[
\begin{array}{cc}
2  &  -1 \\ 1 & 0
\end{array}
\right]_D \left[\begin{array}{c} -1 \\ 1 \end{array} \right]_C=
\left[\begin{array}{c} -3 \\ -1 \end{array} \right].
$$

Moreover, the wrapping numbers are such that $n_A,n_B$ and
$n_{C'}=n_C=N+M_1$ remain invariant, while \bea n_D \rightarrow
n'_D = n_D- (S_C \cdot S_D) n_C = -N+M_2-2M_1 \eea where we see
that we still satisfy the zero charge (anomaly cancellation)
condition, $\sum\limits_i n_i [p_i, q_i]=0$. For the special case
where $N=1,M_1=M_2=0$, we have $n'_D=-1$. The negativity for
$n'_D$ indicates that we should reverse direction of the new
7-brane by changing the cycle into $[-p,-q]$. So finally we obtain
the configuration
$$
A:[1,-1] \quad B:[1,1] \quad D:[1,1] \quad C':[-3,-1]
\qquad
{\rm with}
\quad
n_A=n_B= n_D= n_{C'}=1
$$
Notice that under the above transformation, only the rank of node $D$
changed, so the node $D$ is exactly the node upon which we Seiberg
dualise.
\subsection{Deriving Seiberg Duality on one node from Picard-Lefschetz}
\label{52} First let us recall the rules for Seiberg duality on a
single gauge group (single node) from the point of view of
supersymmetric field theory. For clarification, we consider a
general field theory with only bi-fundamental fields. We use
$a_{ij}>0$ to denote the multiplicity of fields which are
fundamental under $U(n_i)$ and anti-fundamental under $U(n_j)$. In
the quiver diagram this means that there are $a_{ij}$
arrows\footnote{Note that it is possible to have
    $a_{ij}\neq 0$ and $a_{ji}\neq 0$ for given pair $(ij)$.
    This just means that there are arrows from
    $i$ to $j$ as well as arrows from $j$ to $i$.}
starting from node $i$ and ending on node $j$.
The steps of Seiberg duality are:
\begin{itemize}
\item(a) Pick up a node, for example $k$, to do Seiberg duality.

\item(b) Ranks of all other nodes except node $k$ are invariant
  while that of $k$ becomes $N_k-n_k$ where
  $N_k=\sum_{i\neq k}n_i a_{ik}=\sum_{i\neq k}a_{ki} n_i$ is the
  total number of flavors for $U(n_k)$.

\item(c) Reverse the direction of arrows connected to node $k$.
 In field theory, this means that the dual quarks of the gauge group
 $U(N_k-n_k)$
 are in complex conjugate representations to the original quarks in representations of the gauge group
 $U(n_k)$. Therefore $a_{ik}\rightarrow a_{ki},a_{kj}\rightarrow a_{jk}$.

\item(d) Add the Seiberg mesons. If for given $i,j$ we have
 $a_{ik}\neq 0, a_{kj}\neq 0$, there are $m_{ij}=a_{ik}a_{kj}$ arrows
 starting from $i$ to $j$ (if $i=j$ they are adjoint fields). Thus the
 total number of arrows starting from $i$ to $j$ will be $a_{ij}+m_{ij}$.

\item(e) Add the Seiberg superpotential of meson fields and
 dual quarks to the original
 superpotential with the original quarks fields replaced by
 meson fields. If there are fields which acquire mass,
    we simply integrate them out by their equations of motion.

\end{itemize}

Now the issue is how can we explain Seiberg duality from the
geometrical Picard-Lefschetz transformations. Before doing so,
there are a few points which are worth pointing out. First,
Seiberg duality includes action on two parts: the matter part
(quiver diagram) and the superpotential. At this moment, we can
only reproduce the matter part by the geometric Picard-Lefschetz
transformations. It will be interesting to derive the
superpotential from these geometric transformations as
well\footnote{Though in the exceptional collection picture, we can
in principle, though not very conveniently, obtain the
transformation rules for the superpotential as well.}.

Second, even for the matter part, our understanding is not
complete. The reason is that we calculate the quiver diagram by
intersections of cycles in the mirror manifold. The intersection
matrix captures only the antisymmetric part of the quiver diagram,
i.e., we assume that one of $a_{ij}, a_{ji}$ to be zero for any
bi-directional pairs between nodes $i$ and $j$ as explained at the
beginning of Section \ref{five}.
It is important to note that only under these premises can we
derive the matter part of Seiberg duality from
geometric Picard-Lefshets transformation.


Now we show how to reproduce the matter part of Seiberg duality
from geometric Picard-Lefschetz transformations, by comparison of
the quiver duality rules above with those rules at the beginning
of this section. We find that it is important to distinguish other
nodes relative to the node $k$, the dualized node. In the spirit
of \cite{seiberg}, these nodes fall into three categories: the
ones such that only $a_{ik}\neq 0$, the ones such that only
$a_{kj}\neq 0$ and those with both $a_{ik}=a_{ki}=0$. For
simplicity, we call them ``Out'', ``In'' and ``No''  respectively.
Furthermore, we make another important assumption: {\it the order
of cycles relative to cycle $S_{k}$ are
 as $S_{A \in Out}, S_k, S_{B \in In}$ while $S_{p\in No}$ can be
anywhere.}\footnote{recall that $(p,q)$ 7 branes appear with a
natural ordering.} We do not know why this is a necessary
condition, but from the derivation we can see it is indeed
required for Picard-Lefschetz transformation to explain Seiberg
duality. Other ordering would lead to Seiberg-like dualities, but
not the simple Seiberg duality on a single node we are familiar
with. Justifying this condition would be very interesting for the
geometrisation of field theoretic dualities.



Now we proceed with the derivation.
Under this condition of the ordering of cycles,
we move $S_k$ all the way to the left hand side, passing
through all the $S_A$ in the $Out$ category and possibly some
$S_{p}$ in the $No$ category.    We have the following
transformation for each cycle in $A,B,p$:
\beqan
S_A  & \rightarrow &  S_A + (S_A \cdot S_k) S_k\\
S_{k,B,p} & \rightarrow & S_{k,B,p} \\
n_k & \rightarrow & n_k-\sum_{A \in Out} n_A (S_A
\cdot S_k)=n_k-N_k<0 \\
n_{A,B,p}  & \rightarrow & n_{A,B,p}, \eeqan where the sum is
accumulated as we move through each $Out$ cycle. The cycles
$S_{p}$ do not change even if cycle $S_{k}$ has passed them
because they have zero intersection number with  $S_{k}$. Notice
also that the quantity $\sum\limits_A n_A (S_A \cdot S_k)$ is
exactly the number of flavours $N_k$ with respect to the dualising
node $k$. Now since $n_k-N_k<0$, according to our convention, the
rank of the new gauge group should be $\widetilde{n_k}=N_k-n_k$
and the corresponding cycle should be $-S_{k}$. This reproduces
rule (b) of Seiberg duality.

Next we need to calculate the quiver diagram by calculating the
intersection of cycles. First $\widetilde{a_{ik}}=S_i \cdot
(-S_{k})$, which implies that $\widetilde{a_{ik}}=-a_{ik}$. This
explains the reversal of arrows connected to node $k$, i.e., rule
(c). Second, we calculate $\widetilde{a_{ij}}, i, j \neq k$. This
part is modified only when at least one of $i,j$ is in the
category ``Out''. In this case, we have
\begin{eqnarray*}
\widetilde{S_p} \cdot \widetilde{S_A} & = & S_p \cdot S_A \\
\widetilde{S_A} \cdot \widetilde{S_B} & = &  S_A \cdot S_B+
(S_A \cdot S_k) (S_k \cdot S_B) \\
\widetilde{S_{A1}} \cdot \widetilde{S_{A2}} & = &  S_{A1} \cdot
S_{A2}
\end{eqnarray*}
which exactly reproduces rule (d). In summary then we have derived
Seiberg duality from Picard-Lefschetz.

%
\subsection{An Interesting Question}
Now we come to an interesting question. As we saw above, only in
conjunction with the ordering of the cycles and making the special
move of letting node $S_k$ pass through all nodes such that $(S_i
\cdot S_k)>0$, does Picard-Lefschetz monodromy derive Seiberg
duality. Thus indeed the former is a more general class of
phenomenon than the latter. This has been recently pointed out in
\cite{CFIKV} where Seiberg-like dualities were discussed.

If we do not pass through all nodes,
what field theory is given by the transformation? We will see below
that it is not a simple Seiberg Dual theory. It is important to figure out
what is the physics behind such a perfectly well-defined procedure in
geometrical engineering. To demonstrate this point, we continue with
the above example of $\mathbb{F}_{0}$ (see Figure (\ref{PLF0})).
 Starting from Case (I)
$$
\begin{array}{ccccc}
cycle:  & A:[1,-1] &  B:[1,1]  &   C:[-1,1]  & D:[-1,-1]  \\
n:  &  1 & 1 & 1 & 1 \\
S_i \cdot S_D: & -2  &  0  &  2  & 0 \\
\end{array}
$$
Picard-Lefschetz transformation with respect to node D relative to
C, we obtain Case (II)
$$
\begin{array}{ccccc}
cycle:  & A:[1,-1] &  B:[1,1]  &   D:[1,1] & C:[-3,-1] \\
n:  &  1 & 1 & 1 & 1 \\
S_i \cdot S_A: & 0  &  -2  &  -2  & 4 \\
S_i \cdot S_B: & 2  &  0   &  0   & -2 \\
S_i \cdot S_D: & 2  &  0   &  0   & -2 \\
S_i \cdot S_C: & -4 &  2   &  2   & 0 \\
\end{array}
$$
These are the cases discussed earlier and presented in Figure
(\ref{PLF0}).

If we move the node A relative to node C of phase II, we obtain case
(III)
$$
\begin{array}{ccccc}
cycle:  & C:[1,-5] &  B:[1,1]  &  D:[1,1] & A:[-1,1] \\
n:  &  1 & 1 & 1 & 3 \\
S_i \cdot S_A: & -4 &  2  &  2  &  0 \\
S_i \cdot S_B: &  6 &  0  &  0  &  -2 \\
S_i \cdot S_C: &  0 &  -6 &  -6 &  4  \\
S_i \cdot S_D: &  6 &  0  &  0  &  -2 \\
\end{array}
$$
which is Seiberg Dual to (II). However, if we dualise node C
relative to node D of (II), we will obtain Case (IV)
$$
\begin{array}{ccccc}
cycle:  &   A:[1,-1] &  B:[1,1]  & C:[3,1] & D:[-5,-1] \\
n:  &  1 & 1 & 1 & 1 \\
\end{array}
$$
If we further transform node C relative to node B, we get Case (V)
$$
\begin{array}{ccccc}
cycle:  &   A:[1,-1] &  C:[3,1] & B:[-5,-1] & D:[-5,-1] \\
n:  &  1 & 3 & 1 & 1 \\
\end{array}
$$
which is Seiberg dual to (II). We see that after two successive
Picard-Lefschetz moves, we do obtain a Seiberg dual theory. This
hints us that {\it Picard-Lefschetz duality is a fractional
Seiberg-duality}.  The various dualities are summarised in Figure
(\ref{PLF0}).

\onefigure{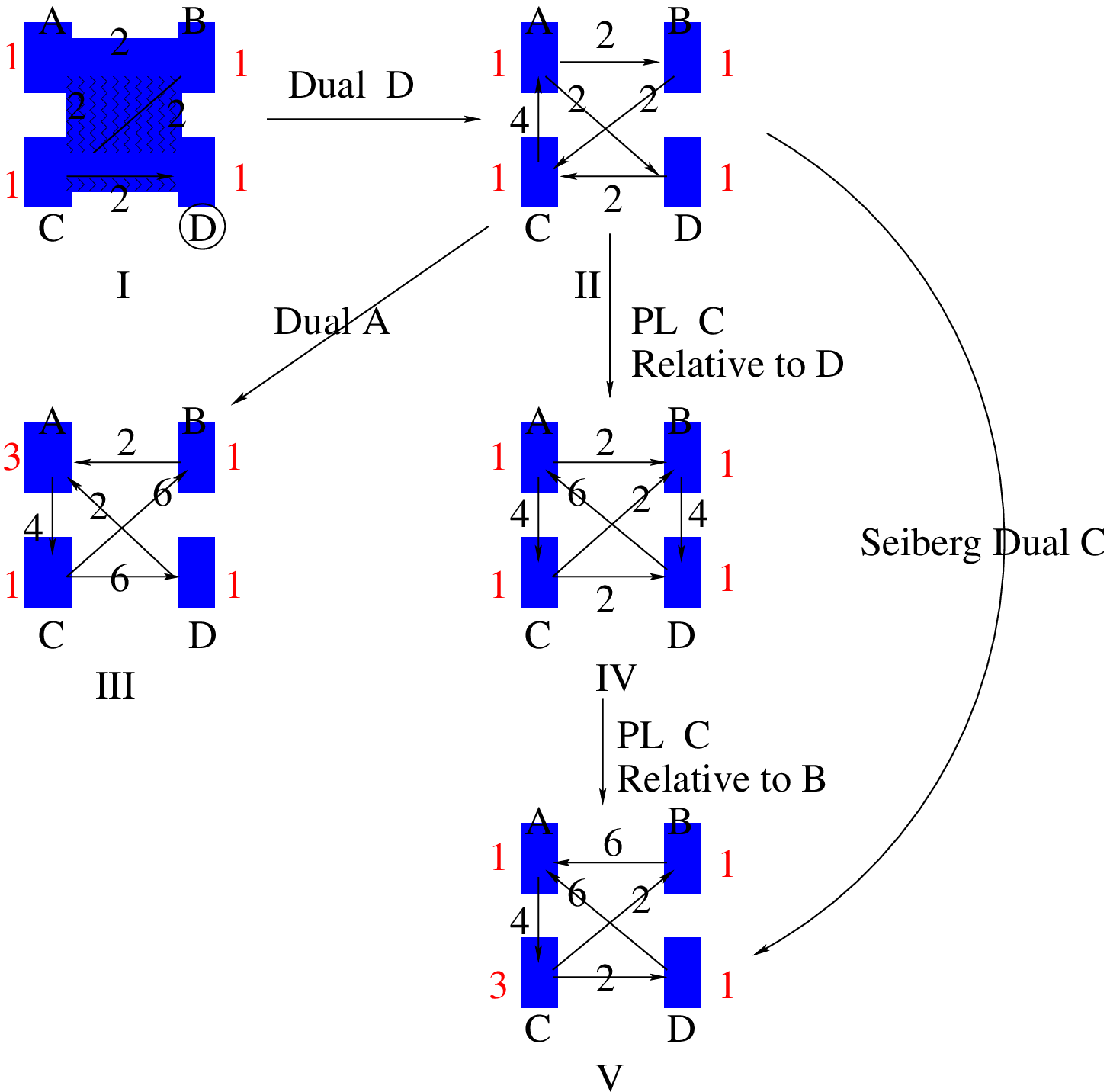}{
The web of dualities one obtains for the zeroth Hirzebruch
    surface $\mathbb{F}_{0}$, as we perform Picard-Lefschetz moves on the
    corresponding $(p,q)$ 7-brane configurations.}
\label{f:PLF0}
\section{Superpotential from Global Symmetries}
\label{glosup}

We have seen that thusfar two alternative methods, one
algebro-geometrical \cite{CFIKV} and another combinatorial
\cite{toric,toric2}, exists in the computation of the matter
content and superpotential. The first problem of finding the
quiver is relatively straight-forward and there exists yet another
prescription using $(p,q)$-brane webs \cite{HaIq}. The
superpotential on the other hand is rather involved: the
$(p,q)$-description so far gives no direct technique, the
exceptional bundle method requires involved Ext computations and
the Inverse algorithm requires nontrivial integrating back.

The problem of finding an efficient method of determining the
superpotential for classes of algebraic singularities remains a
tantalising one. The following observations may yet point us to the
right direction (q.v.~\cite{multi} for discussions in a similar vein,
especially on the third del Pezzo surface).
\subsection{Example: $\mathbb{B}_{0}$}
Let us proceed with the example of the cone over the zeroth del
Pezzo surface, i.e., the blowup of the well-known orbifold
$\IC^3/\IZ_3$ (cf.~page 65 of \cite{CFIKV}). Let us consider phase
II of Figure \ref{f:Sei_dP0}. The 12 fields are arranged as 3 from
node B to A, labelled as $\vec{X}$; 3 from node A to C, labelled
as $\vec{Y}$ and 6 from node C to B, labelled as $Z_{ij}$. Now
$SU(3)$ is the isometry group of $\IC^3/\IZ_3$, thus becomes a
global symmetry group for the gauge theory. We expect the two
triplets of fields ($\vec{X},\vec{Y}$) to be in irreducible
representations of $SU(3)$ and we assign for convenience the
anti-fundamental $\mathbf{\bar{3}}$ of $SU(3)$ while the sextuplet
($Z_{ij}$), to be in the symmetric 6 of $SU(3)$, an invariant
scalar contraction is then obviously $X^i Y^j Z_{ij}$, which is
precisely the superpotential computed by either algebraic methods
or by performing Seiberg duality on phase I of Figure
\ref{f:Sei_dP0}. In general we can follow the tree given in
\cite{CFIKV} modelling all the Seiberg dual theories of the above
and arrive at Equation (\ref{B0}) for all the allowed number of
fields $\mu_{21}$, $\mu_{32}$ and $\mu_{31}$ between nodes $12$,
$23$ and $13$ respectively. (Here to compare to Figure
\ref{f:Sei_dP0} just make the replacements A by 1, B by 2 and C by
3). We expect such numbers to be all solutions for the
ir(reducible) representations of $SU(3)$ and appropriate
contractions then suffice.
\subsubsection{Series of Theories for $\mathbb{B}_{0}$}
The above is but one phase of a series of Seiberg duals of the
theory \cite{CFIKV,seiberg} and we made use of the explicit global
$SU(3)$ flavour symmetry (also see \cite{multi}). We will now show
that using this symmetry alone we can in fact write down the
superpotential uniquely for many phases related to each other by
Seiberg duality.

Let us start from model I in Figure \ref{f:Sei_dP0}. This is the
toric phase \cite{toric}.
In this model, we have 9 fields $X_{AB,i}$,
$X_{BC,i}$ and $X_{CA,i}$ with  $i=1,2,3$, all of which transform
as the fundamental ${\bf 3}$ of the $SU(3)$ flavor symmetry.
There is only one combination (tensor product as a Hom composition)
of these to give a singlet of $SU(3)$, viz.~,
${\bf 3}\otimes {\bf 3}\otimes {\bf 3}={\bf 1}+{\bf 8}+{\bf 8}+{\bf 10}$.
Therefore, the only invariant scalar is given by
\begin{equation}
W_{I} = X_{AB,i}X_{BC,j}X_{CA,k} \epsilon^{ijk}.
\end{equation}
This is of course the well-known superpotential \cite{toric,CFIKV} for
the toric phase of del Pezzo zero.

Now let us perform Seiberg duality with respect to node $A$. This
model II is what was discussed above and in \cite{seiberg,CFIKV}.
We here discuss this example in detail to demonstrate our idea.
First let us analyse the dual quarks. Under Seiberg duality, the
fields $X_{AB,i}$ and $X_{CA,i}$ become $X_{BA}^i$ and $X_{AC}^i$.
Therefore under the $SU(3)$, the ${\bf 3}$ changes to ${\bf
\bar{3}}$. The fields $X_{BC,i}$ are invariant and remain as ${\bf
3}$. Next we need to add the meson fields $M_{CB,ij}$ which should
transform under the tensor product ${\bf 3}\otimes {\bf 3}$. Since
${\bf 3}\otimes {\bf 3}={\bf 6}_{sym} +{\bf\bar{3}}_{antisym}$, we
can write the meson fields into two irreducible representations
$Y_{CB}^{i}$ for ${\bf\bar{3}}$ and $Y_{CB,(ij)}$ for ${\bf 6}$
where $(ij)$ means the symmetrisation of $ij$.

The subsequent superpotential of the dual field theory becomes
$W'=M_{CB,ij} X_{BC,j}\epsilon^{ijk} -M_{CB,ij} X_{BA}^i X_{AC}^i$
where the first term comes from $W_{I}$ and the second term comes
from the duality. Notice that since
$M_{CB,ij}\epsilon^{ijk}=Y_{CB}^{k}$, the first term in $W'$ tells us that
both fields $Y_{CB}^{k}$ and $X_{BC,i}$ are massive and should be
integrated out. Using the equation of motion of fields
$X_{BC,j}$ we find $Y_{CB}^{k}=0$ and the final superpotential is:
\begin{equation}
\label{B0modelII}
W_{II} = -X_{BA}^i X_{AC}^j Y_{CB,(ij)}.
\end{equation}

The above result is derived from applying Seiberg duality rules in
field theory. Now let us show how to use symmetry alone to
reproduce this result. Under Seiberg duality, we have fields which
transform under the following representations of $SU(3)$:
\begin{eqnarray*}
\mbox{Field}  & & \mbox{Rep}(SU(3)) \\
X_{BA}^i & & {\bf \bar{3}}  \\
X_{AC}^i & & {\bf\bar{3}}  \\
X_{BC,i} & & {\bf 3 } \\
M_{CB,ij} & &{\bf 3}\otimes {\bf 3} = {\bf 6}_{sym} +{\bf\bar{3}}_{anti}
    = Y_{CB,(ij)}+ Y_{CB}^{i}
\end{eqnarray*}
Whence we see that the fields $X_{BC,i} ({\bf 3})$ will combine
with fields $Y_{CB}^{i}({\bf\bar{3}})$ to become massive, so they
can be integrated out. The remaining fields are ${\bf
\bar{3}},{\bf\bar{3}}$ and ${\bf 6}$. Symmetry therefore tells us
that there is only one flavor invariant superpotential we can
write down:
$$
W_{II} = X_{BA}^i X_{AC}^j Y_{CB,(ij)},
$$
giving us the same results as (\ref{B0modelII}) with much less work.

Next we dualize with respect to node $C$ to reach model III. In
this case, the meson fields will be ${\bf 6}\otimes {\bf
\bar{3}}:M_{AB,(ij)}^{k}$, which can be decomposed into ${\bf
15}+{\bf 3}$. The ${\bf 3}$ is given by the trace part
$Y_{AB,j}=\sum_{i} M_{AB,(ij)}^{i}$ while the ${\bf 15}$ is given
by the traceless part $Y_{AB,(ij)}^{k}$ with the condition
$\sum_{k}Y_{AB,(kj)}^{k}=0$.  As in model II, the ${\bf 3}$ field
$Y_{AB,j}$ will be integrated out with ${\bf \bar{3}}$ field
$X_{BA}^i$. Thus from these representations we find that the
superpotential is uniquely determined as
$$
W_{III}= Y_{BC}^{(ij)} X_{CA,k} Y_{AB,(ij)}^{k}.
$$

Finally, we dualize on the node $A$ again to reach model IV. It is
the first non-trivial example where the representation is not
irreducible. The meson fields will be $M_{CB,(ij)k}^{l}={\bf
15}\otimes {\bf 3}$ which can be decomposed into ${\bf 24}+{\bf
\overline{15}}+{\bf 6}$. The component of ${\bf 6}$ will become
massive and be integrated out with fields $Y_{BC}^{(ij)}={\bf
\bar{6}}$. This leaves us two irreducible components ${\bf
24}:Y_{CB,(ijk)}^{l}$ with $\sum_{k} Y_{CB,(ijk)}^{k}=0$ and ${\bf
\overline{15}}:Y_{CB,k}^{(ij)}$ with $\sum_k Y_{CB,k}^{(kj)}=0$.
The superpotential is determined again by the $SU(3)$ flavor
symmetry as
$$
W_{IV}=  Y_{BA,l}^{(ik)} X_{AC}^{j}Y_{CB,k}^{(kl)}\epsilon_{ijk}
+Y_{BA,l}^{(ij)} X_{AC}^{k} Y_{CB,(ijk)}^{l}.
$$

We see therefore that by consideration of the representation theory of global
symmetries, one could sometimes obtain the superpotential without recourse to
the Inverse Algorithm or to helix methods.
\onefigure{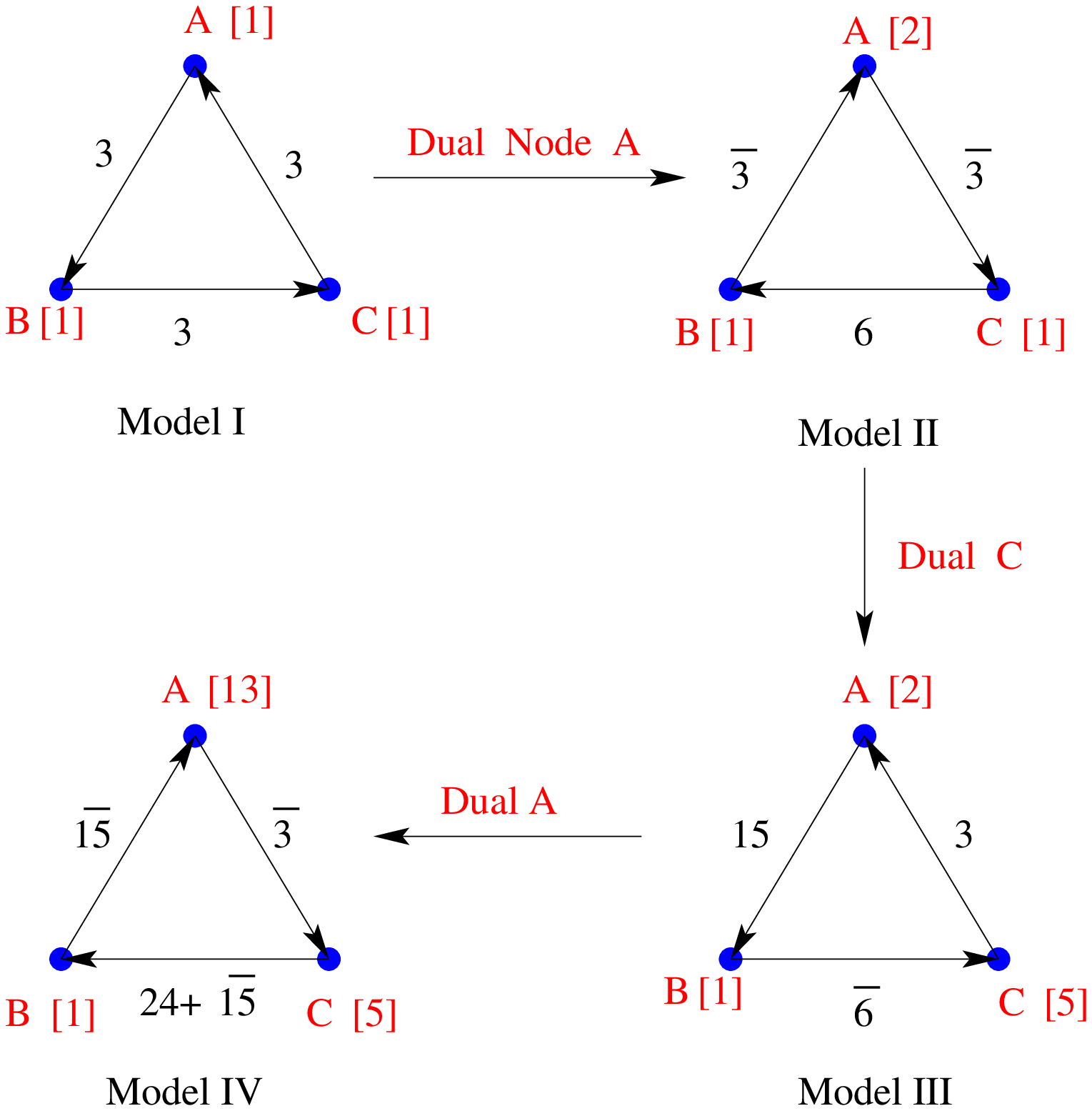}{The first few phases of $\mathbb{B}_0$
obtained by Seiberg duality. The letters A, B, C, label the gauge
group. The numbers in square brackets are the ranks of the gauge
groups.\label{f:Sei_dP0}}
The general prescription seems rather straight-forward, though the
complete justification for this elegant technique still eludes us.
We first identify the isometry of the singularity of concern, and
then group the bi-fundamental fields into irreducible
representations of this symmetry group. Contraction of these
fields, now arranged as tensors of various rank, into a scalar,
should give the final superpotential. Heuristically, this simply
means that there is a remnant global symmetry, perhaps in the form
of the centre of the Lie group, of the enhanced gauge symmetry
which arise in the closed string sector as we compactify Type II
on the appropriate Calabi-Yau cycles. The corresponding closed
string moduli realise as gauge couplings in the open string sector
which lives on the D-brane probe theory, some of which as
coefficients of the terms in the superpotential, whereby giving
our superpotentials surviving symmetries from the geometry.

Mathematically, the superpotential is a sum over all minimal
loops in the quiver, weighted by the dimension of the Ext group of the
various composition of the bundles in the exceptional collection. It
is the non-zero terms that interest us. When the Ext-groups do not
vanish should be precisely determined by the geometry of the
Calabi-Yau base over which we have constructed the bundles.

\subsection{del Pezzo Zero, Markov numbers and Helices}
Let us examine the above case of the del Pezzo zero resolution of
$\IC^3/\IZ_3$ in some more detail. As was pointed out in
\cite{CFIKV}, if we let $(x,y,z)$ denote the number of
bi-fundamentals between the nodes, then one can construct a tree
of branching integer triplets which gives all allowed solutions
under Seiberg duality.

Of course, as introduced earlier in Equation (\ref{B0}) and also
in \cite{CV}, the solutions are dictated by the Diophatine
equation\footnote{For simplicity, we have redefined
$x=\mu_{21},y=\mu_{31},z=\mu_{32}$.}
\begin{equation}
\label{xyz}
x^2 + y^2 + z^2 = x y z,
\end{equation}
which we obtained from tracing over the product of monodromy
matrices. On the other hand if we denote the rank of the nodes to be
$(m,n,p)$, anomaly cancellation demands this triple to be in the
nullspace of the intersection matrix. Whence,
{\tiny $\left(\begin{array}{ccc}
0 & x & -z \cr -x & 0 & y \cr z & -y & 0 \cr
\end{array}
\right)\left(\begin{array}{c} m\\ n\\ p  \end{array}\right)=0$}
and we immediately see the solution $(m,n,p)=a (y,z,x)$ for possibly
fractional $a$ if $(x,y,z)$ were to have a common factor.

We recognise (\ref{xyz}) as a case of the Hurwitz equation
\cite{Hurwitz,Mordell}, the general
solution for which is given in \cite{CV}. One takes the fundamental
solution $(3,3,3)$ and repeatedly applies
{\tiny $\left(\begin{array}{ccc}
-1 & 1 & 1
\cr 0 & 1 & 0 \cr 0 & 0 & 1 \cr
\end{array}
\right)$} in addition to the permutation $\Sigma_3$ on the
triple.\footnote{One should note here that this matrix is nothing
but Seiberg duality on one of the nodes and though discovered
about a century ago was not termed ``Seiberg Duality''.. The notes
here are a gauge theory reinterpretation of these results.} This
generates the tree of solutions. One sees of course that they
generate a braid group, in concordance with the fact that Seiberg
duality is a monodromy action.

With the solution above, we see that the $(x,y,z)$ always have a
common divisor of 3. This means that we can take the $a$ above to be
$1/3$ and obtain the equation
\begin{equation}
\label{3xyz}
m^2 + n^2 + p^2 = 3 m n p
\end{equation}
for the labels of the nodes. We recognize this to be the Markov
equation \cite{Markov}. The solutions for which are the renowned
{\bf Markov numbers} used in Diophantine approximation theory.

Indeed, we see that we are really dealing with only (\ref{3xyz}). Take
(\ref{xyz}), and consider it modolo 3. Because $2^2 \equiv 1^2 (\bmod
3)$, we only have to consider 4 possibilities on the left: $0+0+0$,
$0+0+1$, $0+1+1$ or $1+1+1$. The second and third are instantly discarded
because then the left would be non zero mod 3 while the right divides
3. The fourth is also impossible because the left would be 0 mod 3 and
the right, not so. We conclude that the only solutions to (\ref{xyz})
are when all $(x,y,z)$ are multiples of 3, whereupon we can instantly
rename $(x,y,z) = (3m,3n,3p)$ and obtain (\ref{3xyz}). We summarise:

{\em Theories related by Picard-Lefschetz duality (which are
Seiberg duals) for the world-volume theory of D-branes probing the
cone over del Pezzo zero are characterised by Markov numbers
$(m,n,p)$: it is a $U(m) \times U(n) \times U(p)$ gauge theory
with bifundamental matter {\tiny $\left( \begin{array}{ccc} 0 & 3p
& -3n \cr -3p & 0 & 3m \cr 3n & -3m & 0 \cr \end{array} \right)$}.
}

One might wonder whether our Diophantine equation, derived from the
monodromy matrix condition ${\rm Tr}K = 2$, which is obviously
necessary, is in fact sufficient to describe all solutions. We are
saved by a result of Rudakov \cite{Rudakov1} which proved a 1-1
correspondence between the Fourier-Mukai vector of exceptional bundles
on $\mathbb{P}^2$ related by mutations
(or, in our language, the fractional brane charges on del
Pezzo zero related by monodromy)
and the Markov numbers. Therefore (\ref{3xyz}) does indeed
characterise all solutions.

In a follow-up work \cite{Rudakov2}, Rudakov addressed the
case for $\IF_0$. There, the statements are less powerful than the
$\mathbb{P}^2$  case and a certain subset of the exceptional bundles
are in bijection with $2x^2+y^2+z^2=4xyz$. The
general problem of finding Diophatine equations characterising
exceptional collections over arbitrary varieties remains open.

In fact over the $k$-th del Pezzo surface, the bundles are associated
to the equation of the Markov type as:
\[
ax^2+by^2+cz^2=\sqrt{K^2 abc} \, xyz,
\]
with $a,b,c$ integers and $K^2 = 9 - k$ the intersection number of
the canonical class \cite{Karpov}. The ranks of the what the
authors define to be a triple of ``three-blocks'' of exceptional
collections satisfy the above equation. The precise relation
between this Diophantine equation and the ones discussed in
Section \ref{dPs} eludes us. It could well be that the fact that
they coincide for the simplest case of $\mathbb{P}^2$ is mere
coincidence.
%
%

\section{Conclusions and Prospects}
\label{conclud} We have seen that dualities of quiver theories
become geometric when these theories are realized in the Type IIA
string theory using D6-branes.  Although the superpotential is
difficult to determine in these cases, in some aspects it seems
that D6-brane picture is the more natural one. As we saw in the
del Pezzo cases the geometry of the mirror manifold is able to
provide Diophantine equations completely describing the various
quiver diagrams related to each other by Picard-Lefschetz
transformations. Parenthetically, works on helices on del Pezzo
surfaces have also shown how exceptional collections could be
classified by certain Diophantine equations. We have shown that in
the case of $\mathbb{P}^2$ both prescriptions give the same
Diophantine equation, namely the Markov equation. It would be
enlightening to find out how they are related for the higher del
Pezzos.

Moreover, it would be interesting to see if such equations exist for
the theories arising via orbifolds. It is clear that the equations
represent the necessary and sufficient condition for the existence of
the $T^{3}$ mirror to the zero cycle. For example in the case of
$\mathbb{C}^{3}/\mathbb{Z}_{5}$ orbifold the mirror manifold has five
degenerate fibers of the genus two fibration so we need an equation
which gives a necessary and sufficient condition for the existence of
an eigenvector of the monodromy matrix around five degenerate fibers
of a genus two curve. Such an equation will classify non-compact CY
manifolds with two four cycles and Euler characteristic five similar
to the classification of local del Pezzos from the elliptic curve
\cite{papers}.

As an aside we have shown in detail how with the imposition of
certain condition one could derive Seiberg duality on the matter
content of the quiver theory from Picard-Lefschetz moves; this is
very much in the spirit of \cite{CFIKV}. As an interesting
by-product we have explicited an example where one obtains pairs
of theories as a ``fractional'' generalisation of Seiberg duality.
Of course the full treatment, incorporating the superpotential,
still awaits a geometrical perspective. This should correspond to
interesting behaviour in the field theory and seems to be a
promising direction of pursuit.

Indeed whereas the transformation rules of the matter content
under Seiberg duality are seen as a consequence of
Picard-Lefschetz monodromy, the geometrisation of the
superpotential transformation rules still needs full
understanding. The current methods of computing superpotential,
either from the Inverse Algorithm of \cite{toric,toric2} for toric
singularities, or from the composition of maps of sheafs
\cite{CFIKV}, are computationally intensive. To have something
akin to the elegant rules for the matter content for Seiberg
duality or to determine the terms purely from global symmetries
(isometries of the background geometry) would be a true blessing.

\section*{Acknowledgements}
A.~H.~would like to thank Ron Donagi, and A.~I., Jacques Distler
and Vadim Kaplunovsky for valuable discussions. We would like to
extend our sincere gratitude to the CTP and LNS of MIT and the
Theory Group at UT Austin for their gracious patronage. A.~H.~
would like to thank the organizers of the ``M Theory'' workshop in
the ``Isaac Newton Institute for Mathematical Sciences'' for kind
hospitality during completion of this work. The research of A.~I.~
was suppoted by the NSF under Grant No. 0071512. A.~H.~is also
indebted to the Reed Fund Award and a DOE OJI Award.

\end{document}